\documentclass[aps,prl,twocolumn,floatfix,superscriptaddress]{revtex4-1}
\usepackage{amsfonts,amsmath,amsthm,amssymb,amsopn,graphicx}
\usepackage{dsfont,bbm,xcolor}
\usepackage{enumitem}
\usepackage{hyperref}
\usepackage{breqn}
\usepackage{siunitx}
\usepackage{color}

\makeatletter
\let\cat@comma@active\@empty
\makeatother

\newcommand{\ii}{ {\rm i} }

\newcommand{\ave}[1]{{\langle #1\rangle}}

\def\one{\mathbbm{1}}

\def\one{\mathbbm{1}}

\begin{document}
\title{Self-induced entanglement resonance in a disordered Bose-Fermi mixture}
\author{Juan Jos\'e Mendoza-Arenas}
\affiliation{Departamento de F\'{i}sica, Universidad de los Andes, A. A. 5997 Bogotá, Colombia}
\affiliation{H.H. Wills Physics Laboratory, University of Bristol, Bristol BS8 1TL, UK}
\author{Berislav Bu\v{c}a}
\affiliation{Clarendon Laboratory, University of Oxford, Parks Road, Oxford OX1 3PU, United Kingdom}
\begin{abstract}
Different regimes of entanglement growth under measurement have been demonstrated for quantum many-body systems, with an entangling phase for low measurement rates and a disentangling phase for high rates (quantum Zeno effect). Here we study entanglement growth on a disordered Bose-Fermi mixture with the bosons playing the role of the effective self-induced measurement for the fermions. Due to the interplay between the disorder and a non-Abelian symmetry, the model features an entanglement growth resonance when the boson-fermion interaction strength is varied. With the addition of a magnetic field, the model acquires a dynamical symmetry leading to experimentally measurable long-time local oscillations. At the entanglement growth resonance, we demonstrate the emergence of the cleanest oscillations. Furthermore, we show that this resonance is distinct from both noise enhanced transport and a standard stochastic resonance. Our work paves the way for experimental realizations of self-induced correlated phases in multi-species systems.
%Universality classes of entanglement growth under measurement have been demonstrated for quantum many-body systems with an entangling phase for low measurement rates and a disentangling phase for high measurement rates (Zeno regime) separated by a phase transition \blue{(maybe don't mention phase transition, as this is something we don't present?)}. Here we focus \blue{("study entanglement growth" instead of "focus"?)} on a disordered Bose-Fermi mixture with the bosons playing the role of the effective self-induced measurement for the fermions. Due to the interplay between the disorder and a non-Abelian symmetry we find that the model features a power-law entanglement growth phase \blue{(maybe say is consistent with power law, so the claim on the form is not so strong as we don't know what happens at longer times?)}, separated from the disentangling phase when the boson-fermion interaction strength is varied.  With the addition of a magnetic field, the model acquires a dynamical symmetry leading to experimentally \blue{locally-?}measurable long-time oscillations. At the entanglement growth resonance, which maximises entropy, we demonstrate analytically and numerically \blue{the existence of?} the cleanest oscillations. Furthermore, we show that the entanglement growth resonance is distinct from both \blue{noise} enhanced transport and a standard stochastic resonance. Our work paves the way for experimental realizations of such self-induced phases.
\end{abstract}		

\maketitle
{\it Introduction ---} A crucial question in non-equilibrium physics is how a quantum system thermalizes. This has been understood in recent years in terms of quantum information scrambling \cite{lewis2019nat,landsman2019nat}, i.e. a quantum many-body system, even when isolated from the environment, dynamically encodes quantum information present in the initial state into non-local degrees of freedom by generating entanglement, leaving the local information indistinguishable from that of a thermal state that does not change in time. Such a process maximises entropy. 

Likewise, further maximising entropy by adding external environmental noise to a small system oscillating in time is often believed to decrease the quality of the oscillations. However, stochastic resonance is a fascinating phenomenon whereby an optimal amount of noise can actually increase the strength of a signal \cite{bezrukov1997stochastic} in certain simple quantum or classical systems. In the 90's and 2000's it was intensely studied in a wide variety of biological (e.g. signals in mechanoreceptor neurons) \cite{crayfish} and chemical systems (e.g. reaction-rates) \cite{ChemicalStochasticResonance}, and also for engineering problems (e.g. network optimization) \cite{StochasticResonanceReview}. On the quantum level it has mostly been studied for single (or few) particle systems in an external bath \cite{StochasticResonanceReview,chen2019prb,kato2021njp,wagner2019nat,hussein2020prl}. 

On the other hand, for quantum many body systems only very recently has it been demonstrated that measurements, functioning as an external effective noise, can induce different entanglement entropy growth phases \cite{Li2018prb,skinner2019prx,Ehud1,Ehud2}. These are realized on the level of the individual measurement trajectories and not the ensemble averaged dynamics. 
%Although theoretically very advantageous allowing for both analytical \cite{} and highly efficient numerical \cite{} computations, these quantum circuits and measurement protocols are difficult to implement experimentally. A pressing question is likewise can one find analogous physics that is experimentally more accessible. In particular due to the generic nature of quantum many-body circuits one would expect the main features to persist in different setups. 
Such non-equilibrium physics could be observed experimentally in cold atom systems, which provide a long-standing platform for accessing many-body phenomena under controllable conditions \cite{coldatom1}. Here, the most straightforward measurements are those of local observables.
%\blue{Notably, cold} atom systems provide a long-standing platform for accessing \blue{such} non-equilibrium many-body physics under controllable conditions \cite{coldatom1}. \blue{Here, the} most straightforward measurements are those of local observables. 

It is known that entanglement growth in many-body systems can exhibit long-time oscillations close to criticality \cite{Olalla}. On the other hand, the theory of dynamical symmetries \cite{Buca_2019}, which are a special kind of spectrum generating algebra \cite{Serbyn_2021} fulfilling the principles of extensivity and/or locality, gives the conditions under which persistent oscillations will manifest in local observables. It has been applied to several manifestations of non-stationary dynamics including both closed \cite{Marko1}, Floquet \cite{Chinzei} and dissipative \cite{Buca_2019,BucaJaksch2019,Booker_2020,Carlos1,Jamir1,dissipativeTCobs,Esslinger} time crystals, quantum many-body scars \cite{scarsdynsym1,scarsdynsym2,scarsdynsym3,scarsdynsym4,scarsdynsym5,scarsdynsym6}, attractors \cite{Buca_2020}, synchronization \cite{quantumsynch,buca2021algebraic} and dynamical superconductivity \cite{PhysRevLett.123.030603}. Notably, some systems that have been intensely both studied experimentally \cite{sugawa2011nat,ferrier2014science,delehave2015prl,roy2017prl,lous2018prl,trautmann2018prl} and theoretically \cite{albus2003pra,lewenstein2004prl,pollet2006prl,zujev2008pra,luhmann2008prl,anders2012prl,bukov2014prb,jj2019pra,jj2020pra,jj2021pra}, and that possess dynamical symmetries, are cold-atom Bose-Fermi mixtures, with the bosons playing the role of an emergent bath for the fermions. Although distinct from measurements, such baths have the potential to realise different entanglement growth regimes. More specifically, different disorder realizations play the role of different trajectories, with the coupling between the bosons and fermions functioning as a mutual source of measurement, effectively \emph{self-inducing} such behaviours. This is advantageous in terms of experimental measurements compared to other phenomena such as measurement induced phase transition \cite{skinner2019prx}, dissipative freezing \cite{Carlos1,Carlos2,Kollath} and dark state phase transitions \cite{carollo2021nonequilibrium}, because it does not require indirect observation via e.g. photon counting statistics. 

In this Letter we demonstrate the appearance of an \emph{entanglement growth resonance} in the Boson-Fermion coupling strength for a disordered Bose-Fermi gas in an 1D optical lattice with a dynamical symmetry. We show analytically and numerically that measuring local magnetization oscillations, implied by the dynamical symmetry, it is possible to qualitatively distinguish between different entanglement growth regimes. Our setup, of a standard local quench from a pure state, features the interplay between non-Abelian subsystem symmetries and disorder. As a result, below and above the resonance, where entanglement grows slowly, the oscillations at the single realization level are largely irregular. On the other hand, at the resonance, where the growth is maximal, the cleanest oscillations emerge. Thus one may readily distinguish between these different entangling dynamics without the need for measuring entanglement or post-selection.
%Our setup, of a standard local quench from a pure state, leads to a novel type of entanglement growth behaviour due to the interplay between the non-Abelian subsystem symmetries and disorder. More specifically, below the entanglement growth resonance we find power-law entanglement growth \blue{(this also seems to be the case above the resonance, but for UBF not so big. Also, given the times we reach, seems a claim that it too strong)}, and \blue{well} above it we find, in accordance with quantum Zeno effect, saturation to constant values. At the resonance the entanglement entropy growth is maximal. Our main result is that, due to the dynamical symmetry, by measuring the oscillations of the transverse magnetization of the fermions, one may readily distinguish between these phases without the need for measuring entanglement or post-selection. Moreover and counter-intuitively, the cleanest oscillations occur for maximal entropy growth, which we demonstrate both analytically and numerically. \blue{It seems to me that the claim that we can distinguish entangling and disentagling phases from the oscillations is too strong. The oscillations are very clean at the resonance, but below and above it they are not, and we could not tell what is below and what is above.} 
We finally show that the entanglement growth resonance is distinct from both well-known stochastic resonances and noise enhanced transport \cite{Plenio_2008,Rebentrost_2009,jj2013prb,Contreras_Pulido_2014,gorman2018prx,potocnik2018nat,maier2019prl,elinor2020prr}, signalling a novel behaviour of quantum matter.

\begin{figure}[t]
\centering
\includegraphics[scale=0.63]{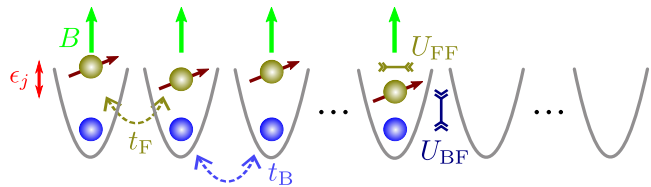}
\caption{Scheme of the Bose-Fermi mixture showing entanglement growth resonance. The blue circles represent the hard-core bosons, which hop between neighboring sites. The orange circles represent the fermions with spin initially projected along the $x$ axis. These hop between nearest neighbors, repel each other and the bosons on the same site, experience an external magnetic field and disorder. Initially, only the first half of the lattice is occupied.}
\label{model_figure}
\end{figure}

{\it Model and setup ---}
Our proposal of establishing an entanglement growth resonance is based on a mixture of hard-core bosons and spin-$1/2$ fermions, depicted in Fig. \ref{model_figure} and described by the one-dimensional Bose-Fermi-Hubbard model
\begin{align} \label{bose_fermi_hami}
    H_{\text{BF}}=H_{\text{B}}+H_{\text{F}}+U_{\text{BF}}\sum_{j}n_j^{\text{B}}\left(n_{j,\uparrow}^{\text{F}}+n_{j,\downarrow}^{\text{F}}\right).
\end{align}
Here $H_{\text{B}}$ is the bosonic Hamiltonian, given by
\begin{align} \label{bose_hami}
    H_{\text{B}}=-t_{\text{B}}\sum_{\langle i,j\rangle}\left(b_i^{\dagger}b_j+\text{H.c.}\right),
\end{align}
and $H_{\text{F}}$ is the fermionic component, corresponding to
\begin{align} \label{fermi_hami}
\begin{split}
    H_{\text{F}}&=-t_{\text{F}}\sum_{\langle i,j\rangle,\sigma}\left(f_{i,\sigma}^{\dagger}f_{j,\sigma}+\text{H.c.}\right)\\
    &+\sum_{j}\left[U_{\text{FF}}\ n_{j,\uparrow}^{\text{F}}n_{j,\downarrow}^{\text{F}}+\epsilon_jn_j^{\text{F}}+\frac{B}{2}\left(n_{j,\uparrow}^{\text{F}}-n_{j,\downarrow}^{\text{F}}\right)\right].
    \end{split}
\end{align}
We consider a lattice of $L$ sites. Operator $b_j^{\dagger}$ ($b_j$) creates (annihilates) a boson on site $j$, $n_j^{\text{B}}=b_j^{\dagger}b_j$ is the local boson number operator, $f_{j,\sigma}^{\dagger}$ ($f_{j,\sigma}$) creates (annihilates) a fermion with spin $\sigma=\uparrow,\downarrow$ on site $j$, and $n^{\text{F}}_{j,\sigma}=f_{j,\sigma}^{\dagger}f_{j,\sigma}$ is the local fermion number operator for spin $\sigma$. We also define $n^{\text{F}}_{j}=n^{\text{F}}_{j,\uparrow}+n^{\text{F}}_{j,\downarrow}$. The bosonic Hamiltonian in Eq. \eqref{bose_hami} corresponds to nearest-neighbor ($\langle i,j\rangle$) hopping processes with amplitude $t_{\text{B}}$; there are no on-site interactions due to the hard-core nature of the bosons. The fermionic Hamiltonian in Eq. \eqref{fermi_hami} incorporates nearest-neighbor hopping with amplitude $t_{\text{F}}$, on-site interaction $U_{\text{FF}}$, magnetic field $B$ and site-dependent on-site potential $\epsilon_j$ selected as a uniformly-distributed random number in the interval $[0,h]$. Finally, $U_{\text{BF}}$ in Eq. \eqref{bose_fermi_hami} is the boson-fermion coupling, which gives the amplitude of the noise on the fermions due to the bosons and vice versa. In this form each species serves as a non-Markovian bath to the other, which we consider explicitly rather than tracing one of them out. We note that such a system can be implemented with a mixture of $^{174}$Yb, which is a bosonic isotope with zero nuclear spin, and $^{171}$Yb, being a fermionic isotope with nuclear spin $I=1/2$ \cite{YTakasu-JPSJ09}. Furthermore, recent experimental advances might be exploited to engineer systems where only one species experiences disorder \cite{rubio2019prx}.   

To assess the entanglement growth resonance phenomenon, we calculate the dynamics of the system under a quench. The initial product state is given by a fermion and a boson on sites $j=1,...,L/2$, with the second half of the lattice being empty. In addition, the spin of each fermion is initially projected along the $x$ axis, thus being on a state $1/\sqrt{2}\left(f^{\dagger}_{\downarrow}+f^{\dagger}_{\uparrow}\right)|0\rangle$. We set the energy scale by taking $t_{\text{B}}=t_{\text{F}}$=1, and consider interacting fermions with $U_{\text{FF}}=1$, $B=16$, $h=8$ and several values of $U_{\text{BF}}$. We simulate the time evolution for $L=24$, a final time $T=12$ and up to 50 disorder realizations using the time-evolving block decimation, describing the system as a matrix product state (MPS) \cite{vidal2004prl,paeckel2019ann}, which is largely applied to boson-fermion mixtures \cite{jj2019prb,stolpp2020arxiv}. Our calculations were performed using the open-source Tensor Network Theory library \footnote{Calculations were carried out using an optimized version of the Tensor Network Theory library \cite{tnt,tnt_review1} being developed by Paul Secular.}. 

{\it Entanglement growth in a Bose-Fermi mixture ---} The model has a dynamical symmetry. The disordered potential and the fermion-boson coupling break the usual $SU(2)$ $\eta$-pairing symmetry of the system \cite{essler_frahm_gohmann_klumper_korepin_2005} leaving only the spin $SU(2)$ symmetry,
\begin{equation}
    [H_{\text{F}},S^z] = 0, \ [H_{\text{F}},S^\pm] = \pm BS^\pm,    
\end{equation}
where $S^\alpha=\sum_{j=}^L S^\alpha_j$, $S^z_j = \left[n^{\text{F}}_{j, \uparrow}-n^{\text{F}}_{j,\downarrow}\right]$, $S^+_j = f_{j, \uparrow}^\dag f_{j, \downarrow}$ and $S^-_j =  f_{j, \downarrow}^\dag f_{j, \uparrow}$. The dynamical symmetry is based on this trivial $SU(2)$ spin symmetry. However, it has non-trivial implications in the presence of the disorder and coupling to bosons.
More specifically,
\begin{equation}
[H,A_s \otimes \one_b]=B A_s \otimes \one_b,
\end{equation}
where $A_s=S^+$ and $\one_b$ is the boson identity operator. As this dynamical symmetry fulfills the trivial $SU(2)$ algebra, we may eliminate the magnetic field by means of transformation to a co-rotating basis $U(t)=\exp[\ii B S^z t]$, and obtain
\begin{equation}
H_{\text{rot}}=U(t)H_{\text{BF}}U(t)^\dagger=H_{\text{BF}}-\frac{B}{2}S^z,
\end{equation}
i.e. $[H_{\text{rot}},A_s\otimes\one]=[H_{\text{rot}},A^\dagger_s\otimes\one]=0$. Then invoking thermalization arguments \cite{BenjaminGGE} we know that, following a generic quench, the time evolution in the rotating frame will drive the system into an effective maximal entropy state. This state is fully determined by the generalized temperatures given by the initial expectation values of the only conserved extensive operators, namely $\ave{H_{\text{rot}}} ,\ave{S^z},\ave{S^\pm},\ave{N^{\text{F}}}$ and $\ave{N^{\text{B}}}$, where $N^{\text{F/B}}=\sum_j n_j^{\text{F/B}}$ is the total fermion/boson number. In particular, generalised temperatures $\mu^{\pm}$ correspond to the expectation values $\ave{S^\pm}$, where $\mu^{+}=(\mu^{-})^{*}$ due to hermiticity. Transforming back gives that $\mu^{\pm}$ become time-dependent with frequency $B$, namely $\mu^{\pm} \to \mu^{\pm} e^{\pm \ii B t}$. That is, in the laboratory frame, the only long-time oscillations take place at frequency $\omega=B$ for local observables $O$ that have non-zero overlap with $S^\pm$, i.e. $\ave{S^+ O} \neq 0$ \cite{Buca_2019,Marko1}. The dynamics will proceed generically through scrambling up to the (dynamical) symmetries. Therefore, the largest entropy will be associated to oscillations at frequency $\omega=B$ only. The transient dynamics, before the system has reached maximum entropy, can feature oscillations at more frequencies. This immediately implies that maximum entropy gives the cleanest oscillations. 
\begin{figure}[t]
\begin{center}
\includegraphics[scale=0.75]{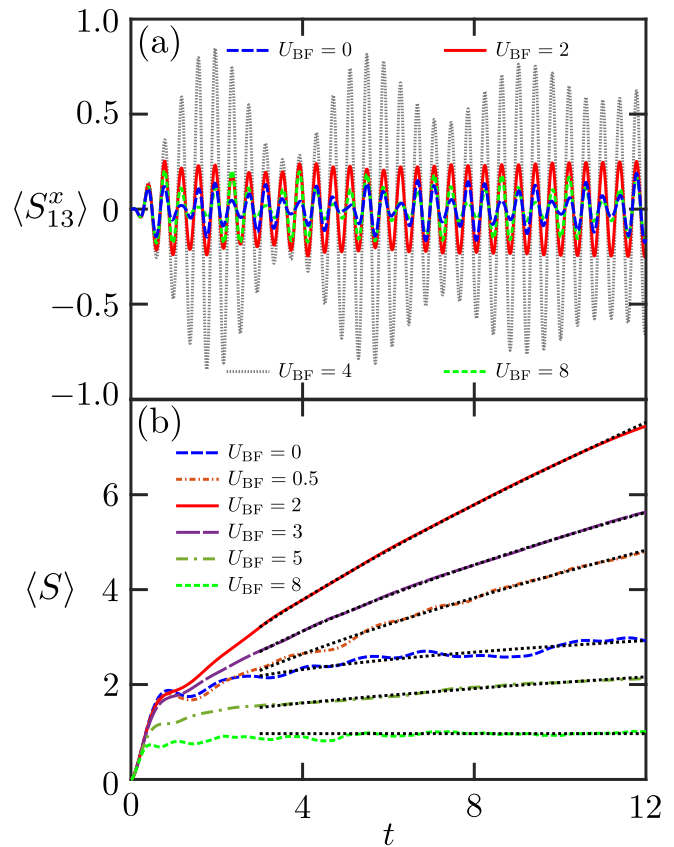}
\end{center}
\caption{(a) Spin oscillations on site $\frac{L}{2}+1$ along the $x$ axis for the same disorder realization and different values of $U_{BF}$. (b) Disorder-averaged von Neumann entropy $\langle S\rangle$ as a function of time for the central bipartition and different values of $U_{\text{BF}}$. The dotted black lines correspond to power law fits, with $\alpha=0.23(12),0.68(2),0.732(3),0.562(4),0.26(4),0$ for $U_{\text{BF}}=0,0.5,2,3,5,8$.}
%\blue{Improve lower panel for 0 and 0.5, leave same colors for 0 and 2 in upper and lower figures. For guide to the eye, include constant for large $U_{BF}$. Include $U_{BF}=8$.}
%Considering \cite{prelovsek2016prb}, according to which our parameters correspond to nonergodic charge but ergodic spin dynamics, entanglement should grow as power law for $U_{BF}=0$ (no bosons) instead of the expected log behavior for true MBL. The results are not soft enough for a good fit, but the power law gives an exponent of $0.20$, close to the $0.18$ of \cite{prelovsek2016prb}.}
\label{resonance_oscil_site13}
\end{figure}

To illustrate this, we simulate the time evolution of the initial state for several values of the boson-fermion coupling $U_{\text{BF}}$, and focus on fermionic observables. Namely, in Fig. \ref{resonance_oscil_site13}(a) we show the magnetization along the $x$ direction $\langle S_j^x\rangle=\langle S_j^++S_j^-\rangle$ on the first initially-empty site ($j=13$) as a function of time for a few $U_{\text{BF}}$ values and the same disorder realization. In the absence of boson-fermion coupling, due to the disorder being only in one of the $SU(2)$ symmetry sectors, the fermions are not fully localized and a finite but slow transport is expected \cite{prelovsek2016prb}. 
%\blue{(do we have the same type of separated dynamics here? Spin is not transported if charge isn't?)}. 
This is appreciated in the corresponding low magnetization shown in Fig. \ref{resonance_oscil_site13}(a). On the other hand, it is widely appreciated that external noise can enhance transport in such situations \cite{vznidarivc2017dephasing}, and even melt full localization in other models \cite{luschen2017prx,gopalakrishnan2017prl}. In our case, such effect is provided by the bosons, which do not experience disorder. Indeed, turning on the boson-fermion coupling initially enhances the propagation of fermions (see below), leading to a larger magnetization, as seen in Fig. \ref{resonance_oscil_site13}(a). Notably, the oscillations emerging from the dynamical symmetries of the model, which result in non-stationarity \cite{Buca_2019,PhysRevLett.123.030603}, feature the cleanest pattern (in the sense of the most homogeneous amplitude) for an intermediate value of $U_{\text{BF}}$, here $U_{\text{BF}}=2$. Stronger boson-fermion couplings might lead to larger amplitudes, but the oscillations become more irregular; even larger coupling, which results in slow-moving interspecies composite quasiparticles and thus decreases the magnetization signal (i.e. a corresponding quantum Zeno effect), leads to highly uneven oscillations. The same behavior is seen for the other initially-empty sites of the lattice, and crucially, for the different simulated disorder realizations. Thus, at the single-realization level, an intermediate noise induces the cleanest signal. 

Remarkably, this boson-fermion coupling also results in maximal entanglement across the system. We show this by calculating the time evolution of the von Neumann entropy $S$ of the bipartition corresponding to the initially occupied and empty lattice sites, readily obtained from the numerical method. This is defined as 
\begin{equation}
    S=-\text{tr}\left(\rho_{o}\log_2\rho_{o}\right)=-\sum_{\alpha}\lambda_{\alpha}^2\log_2\lambda_{\alpha}^2,
\end{equation}
with $\rho_{\text{o}}$ the reduced density matrix of the initially-occupied half of the system, and $\lambda_{\alpha}^2$ its eigenvalues. Note that $S$ incorporates the quantum correlations of both fermions and bosons, and is directly related to the size of the MPS describing the system.  

\begin{figure}[t]
\begin{center}
\includegraphics[scale=0.75]{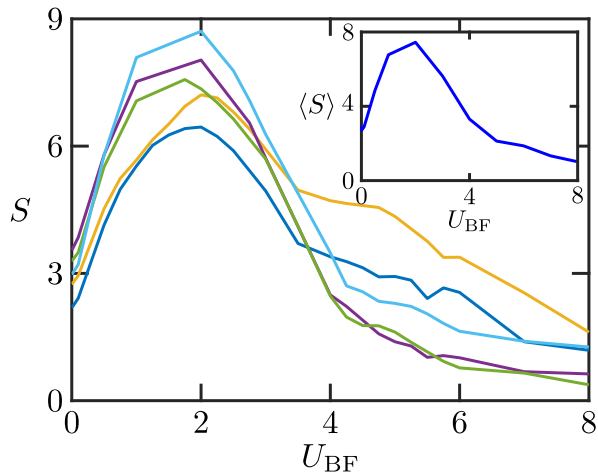}
\end{center}
\caption{Final von Neumann entropy at the central bipartition as a function of $U_{\text{BF}}$. Main panel: Examples of single realizations, where the same disorder distribution is used for each curve. Inset: Average over all disorder realizations.}
%\blue{For one of the realizations the maximum is slightly below 2. Is the maximal entanglement also there?}
\label{entang_figure}
\end{figure}

The dynamics of the von Neumann entropy, averaged over several realizations, is shown in Fig. \ref{resonance_oscil_site13}(b). In agreement with the argument discussed above, the fastest entanglement growth takes place at $U_{\text{BF}}=2$. Moreover, up to the final simulated time, the evolution is consistent with a power law, $S\propto t^\alpha$, with $\alpha<1$ being maximal at the same boson-fermion coupling. Also, for $U_{\text{BF}}\gg1$ we find saturation to an area law behavior ($S\propto t^0$). Importantly, this is true also at the single realization level. We exemplify this in Fig. \ref{entang_figure} for five different disorder realizations, and see that the final entropy is maximal at (or very close to) $U_{\text{BF}}=2$, a behavior that translates directly into the average (see inset). For these realizations, we checked that the cleanest oscillations also occurred at the same value of $U_{\text{BF}}$. Our observations thus suggest a strategy to determine conditions that optimize entanglement by performing measurements of single-site observables. 
%We also note that in the absence of interactions between fermions \blue{... (see SM or not shown), evidencing that our main result is truly a many-body effect}.    

\begin{figure}[t]
\begin{center}
\includegraphics[scale=0.75]{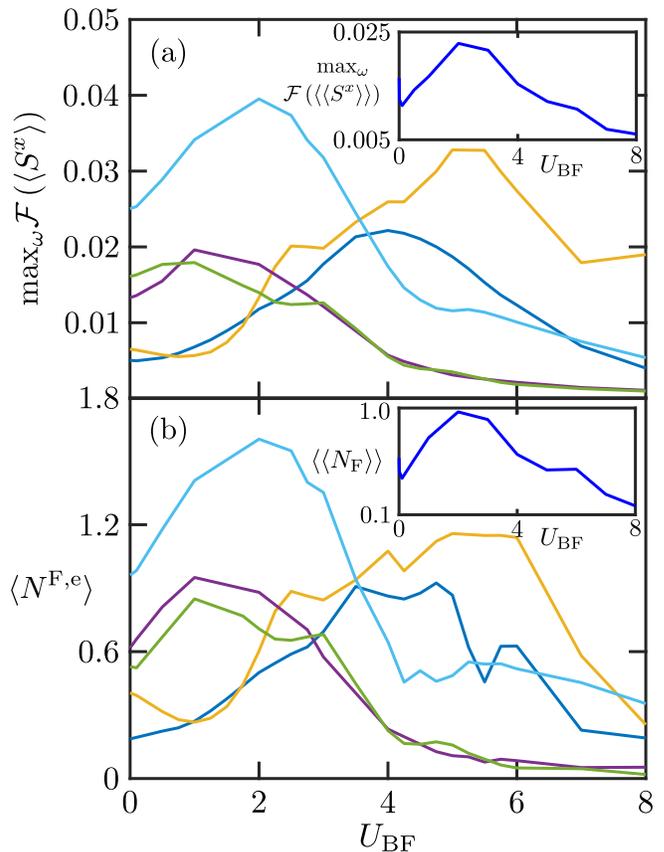}
\end{center}
\caption{(a) Maximum of Fourier transform of magnetization oscillations in $x$ averaged over the second half of the lattice, as a function of $U_{BF}$, for the same disorder realizations of Fig. \ref{entang_figure}. (b) Final population of fermions in the second half of the lattice, as a function of $U_{BF}$, for the same disorder realizations of (a). Insets: Averages over all disorder realizations.}
\label{stoch_res_figure}
\end{figure}

We emphasize that the unravelled phenomenon is fundamentally different from a stochastic resonance originated from the enhancement of fermionic transport due to the noise caused by the bosons. To illustrate this, we depict in Fig. \ref{stoch_res_figure}(a) the maximum of the Fourier transform of the $x$ magnetization averaged over the initially-empty half of the system, $\langle S^x\rangle=\sum_{j=\frac{L}{2}+1}^L \langle S_j^x\rangle$, for the same five disorder realizations of Fig. \ref{entang_figure}, and in Fig. \ref{stoch_res_figure}(b) we show the final number of fermions in the same subsystem, $\langle N^{\text{F,e}}\rangle=\sum_{j=\frac{L}{2}+1}^L \langle n_j^{\text{F}}\rangle$; the insets correspond to the averages over all the disorder realizations. Even though the latter present a maximum at $U_{\text{BF}}=2$, this does not hold at the single disorder realization level. For each one, the optimal transport of fermions coincides with the highest Fourier peak of the magnetization oscillations. This indicates that the stochastic resonance (largest oscillations) results when the fermions maximally populate the second half of the system, and that it manifests without any external bath but rather is emergent by the many-body system itself. However, it can occur at weaker or stronger boson-fermion couplings than that of the entanglement growth resonance, depending on the particular details of the realization; the mechanism of both resonances is thus different. 

Finally, a few important points are in order. On the one hand, the dynamics of the bosons is quite different to that of the fermions. They do not feature a resonance as in Fig. \ref{stoch_res_figure}(b); instead their population on the initially-empty half of the system (at a given time) generally decays monotonically with $U_{\text{BF}}$ (not shown). For very large boson-fermion couplings, they get localized along with the fermions, as a result of heavy quasiparticle formation; this could be pictured as a quantum Zeno mechanism of the fermions on the bosons. Also, for a system without disorder, the non-Abelian symmetry alone does not induce a clean entanglement resonance. 
%Further effects of systems on baths of a small number of degrees of freedom, such as proximity effects \cite{nandkishore2015prb,hyatt2017prb,rubio2019prx}, could be explored with similar multi-species setups. 
%\blue{We also note that for a clean system without disorder, the non-Abelian symmetry alone does not induce a clean entanglement resonance and the growth seems to be qualitatively similar up to large values of $U_{BF}$ where we see Zeno-induced saturation Fig.}

{\it Conclusion ---} Based on the theory of dynamical symmetries, we have shown that a prototypical model of a Bose-Fermi mixture in an optical lattice features an entanglement growth resonance, which presents the cleanest persistent oscillations of the transverse magnetization in one half of the lattice following a local quench. Remarkably, since such oscillations correspond to single-site observables, this motivates the design of protocols to maximise entanglement in many-body systems guided by local measurements.

The origin of this phenomenon is fundamentally different from the usual single-particle non-extensive stochastic resonance \cite{StochasticResonanceReview}. Firstly, it occurs in an extensive quantum many-body system. Secondly, it occurs without introducing an external bath at some temperature; our many-body model functions as its own bath emerging from a pure initial state quench, self-inducing the resonance. Finally, it is based on a complex interplay of disorder, dynamical symmetries, quantum Zeno effect and noise-enhanced transport, being beyond a stochastic resonance mechanism caused solely by the latter.
%Our work therefore lifts the phenomena of stochastic resonance to a fundamental one in extended quantum many-body systems. 

Our results motivate the exploration of the interplay between dynamical symmetries and other types of non-Markovian baths. Furthermore, they inspire future research on connections between entanglement growth and transport properties of multi-species systems with dynamical symmetries. Isolated disordered fermionic lattices already feature nontrivial relaxation when the disorder is coupled to a particular type of degrees of freedom \cite{prelovsek2016prb}, and their interaction with clean baths of a small number of particles leads to even richer physics such as proximity effects \cite{nandkishore2015prb,hyatt2017prb,rubio2019prx}. In addition, our setup could be exploited to design novel mechanisms of dynamical superfluidity in Bose-Fermi mixtures \cite{PhysRevLett.123.030603}.

{\it Acknowledgements ---} J. J. M.-A. gratefully acknowledges support from Ministerio de Ciencia, Tecnolog\'ia e Innovaci\'on (MINCIENCIAS), through the project Producci\'on y Caracterizaci\'on de Nuevos Materiales Cu\'anticos de Baja Dimensionalidad: Criticalidad Cu\'antica y Transiciones de Fase Electr\'onicas (Grant No. 120480863414), from Fundaci\'on para la Promoci\'on de la Investigaci\'on y la Tecnolog\'ia of Banco de la Rep\'ublica, through the project Control de ondas de densidad de carga y superconductividad en sistemas fermi\'onicos por medio de forzamiento peri\'odico (Grant No. 4308), and from the UK's Engineering and Physical Sciences Research Council (EPSRC) under grant EP/T028424/1. BB gratefully acknowledges funding from EPSRC programme grant EP/P009565/1, EPSRC National Quantum Technology Hub in Networked Quantum Information Technology (EP/M013243/1), and the European Research Council under the European Union's Seventh Framework Programme (FP7/2007-2013)/ERC Grant Agreement no. 319286, Q-MAC.

\bibliography{main}

\begin{thebibliography}{78}%
\makeatletter
\providecommand \@ifxundefined [1]{%
 \@ifx{#1\undefined}
}%
\providecommand \@ifnum [1]{%
 \ifnum #1\expandafter \@firstoftwo
 \else \expandafter \@secondoftwo
 \fi
}%
\providecommand \@ifx [1]{%
 \ifx #1\expandafter \@firstoftwo
 \else \expandafter \@secondoftwo
 \fi
}%
\providecommand \natexlab [1]{#1}%
\providecommand \enquote  [1]{``#1''}%
\providecommand \bibnamefont  [1]{#1}%
\providecommand \bibfnamefont [1]{#1}%
\providecommand \citenamefont [1]{#1}%
\providecommand \href@noop [0]{\@secondoftwo}%
\providecommand \href [0]{\begingroup \@sanitize@url \@href}%
\providecommand \@href[1]{\@@startlink{#1}\@@href}%
\providecommand \@@href[1]{\endgroup#1\@@endlink}%
\providecommand \@sanitize@url [0]{\catcode `\\12\catcode `\$12\catcode
  `\&12\catcode `\#12\catcode `\^12\catcode `\_12\catcode `\%12\relax}%
\providecommand \@@startlink[1]{}%
\providecommand \@@endlink[0]{}%
\providecommand \url  [0]{\begingroup\@sanitize@url \@url }%
\providecommand \@url [1]{\endgroup\@href {#1}{\urlprefix }}%
\providecommand \urlprefix  [0]{URL }%
\providecommand \Eprint [0]{\href }%
\providecommand \doibase [0]{http://dx.doi.org/}%
\providecommand \selectlanguage [0]{\@gobble}%
\providecommand \bibinfo  [0]{\@secondoftwo}%
\providecommand \bibfield  [0]{\@secondoftwo}%
\providecommand \translation [1]{[#1]}%
\providecommand \BibitemOpen [0]{}%
\providecommand \bibitemStop [0]{}%
\providecommand \bibitemNoStop [0]{.\EOS\space}%
\providecommand \EOS [0]{\spacefactor3000\relax}%
\providecommand \BibitemShut  [1]{\csname bibitem#1\endcsname}%
\let\auto@bib@innerbib\@empty
%</preamble>
\bibitem [{\citenamefont {Lewis-Swan}\ \emph {et~al.}(2019)\citenamefont
  {Lewis-Swan}, \citenamefont {Safavi-Naini}, \citenamefont {Bollinger},\ and\
  \citenamefont {Rey}}]{lewis2019nat}%
  \BibitemOpen
  \bibfield  {author} {\bibinfo {author} {\bibfnamefont {R.~J.}\ \bibnamefont
  {Lewis-Swan}}, \bibinfo {author} {\bibfnamefont {A.}~\bibnamefont
  {Safavi-Naini}}, \bibinfo {author} {\bibfnamefont {J.~J.}\ \bibnamefont
  {Bollinger}}, \ and\ \bibinfo {author} {\bibfnamefont {A.~M.}\ \bibnamefont
  {Rey}},\ }\href {\doibase 10.1038/s41467-019-09436-y} {\bibfield  {journal}
  {\bibinfo  {journal} {Nat. Commun.}\ }\textbf {\bibinfo {volume} {10}},\
  \bibinfo {pages} {1581} (\bibinfo {year} {2019})}\BibitemShut {NoStop}%
\bibitem [{\citenamefont {A.}\ \emph {et~al.}(2019)\citenamefont {A.},
  \citenamefont {Figgatt}, \citenamefont {Schuster}, \citenamefont {Linke},
  \citenamefont {Yoshida}, \citenamefont {Yao},\ and\ \citenamefont
  {Monroe}}]{landsman2019nat}%
  \BibitemOpen
  \bibfield  {author} {\bibinfo {author} {\bibfnamefont {L.~K.}\ \bibnamefont
  {A.}}, \bibinfo {author} {\bibfnamefont {C.}~\bibnamefont {Figgatt}},
  \bibinfo {author} {\bibfnamefont {T.}~\bibnamefont {Schuster}}, \bibinfo
  {author} {\bibfnamefont {N.~M.}\ \bibnamefont {Linke}}, \bibinfo {author}
  {\bibfnamefont {B.}~\bibnamefont {Yoshida}}, \bibinfo {author} {\bibfnamefont
  {N.~Y.}\ \bibnamefont {Yao}}, \ and\ \bibinfo {author} {\bibfnamefont
  {C.}~\bibnamefont {Monroe}},\ }\href {\doibase 10.1038/s41586-019-0952-6}
  {\bibfield  {journal} {\bibinfo  {journal} {Nature}\ }\textbf {\bibinfo
  {volume} {567}},\ \bibinfo {pages} {61} (\bibinfo {year} {2019})}\BibitemShut
  {NoStop}%
\bibitem [{\citenamefont {Bezrukov}\ and\ \citenamefont
  {Vodyanoy}(1997)}]{bezrukov1997stochastic}%
  \BibitemOpen
  \bibfield  {author} {\bibinfo {author} {\bibfnamefont {S.~M.}\ \bibnamefont
  {Bezrukov}}\ and\ \bibinfo {author} {\bibfnamefont {I.}~\bibnamefont
  {Vodyanoy}},\ }\href@noop {} {\bibfield  {journal} {\bibinfo  {journal}
  {Nature}\ }\textbf {\bibinfo {volume} {385}},\ \bibinfo {pages} {319}
  (\bibinfo {year} {1997})}\BibitemShut {NoStop}%
\bibitem [{\citenamefont {Douglass}\ \emph {et~al.}(1993)\citenamefont
  {Douglass}, \citenamefont {Wilkens}, \citenamefont {Pantazelou},\ and\
  \citenamefont {Moss}}]{crayfish}%
  \BibitemOpen
  \bibfield  {author} {\bibinfo {author} {\bibfnamefont {J.~K.}\ \bibnamefont
  {Douglass}}, \bibinfo {author} {\bibfnamefont {L.}~\bibnamefont {Wilkens}},
  \bibinfo {author} {\bibfnamefont {E.}~\bibnamefont {Pantazelou}}, \ and\
  \bibinfo {author} {\bibfnamefont {F.}~\bibnamefont {Moss}},\ }\href {\doibase
  10.1038/365337a0} {\bibfield  {journal} {\bibinfo  {journal} {Nature}\
  }\textbf {\bibinfo {volume} {365}},\ \bibinfo {pages} {337} (\bibinfo {year}
  {1993})}\BibitemShut {NoStop}%
\bibitem [{\citenamefont {Zhu}\ and\ \citenamefont
  {Shu~Li}(2002)}]{ChemicalStochasticResonance}%
  \BibitemOpen
  \bibfield  {author} {\bibinfo {author} {\bibfnamefont {R.}~\bibnamefont
  {Zhu}}\ and\ \bibinfo {author} {\bibfnamefont {Q.}~\bibnamefont {Shu~Li}},\
  }\href {\doibase 10.1039/B106072B} {\bibfield  {journal} {\bibinfo  {journal}
  {Phys. Chem. Chem. Phys.}\ }\textbf {\bibinfo {volume} {4}},\ \bibinfo
  {pages} {82} (\bibinfo {year} {2002})}\BibitemShut {NoStop}%
\bibitem [{\citenamefont {Gammaitoni}\ \emph {et~al.}(1998)\citenamefont
  {Gammaitoni}, \citenamefont {H\"anggi}, \citenamefont {Jung},\ and\
  \citenamefont {Marchesoni}}]{StochasticResonanceReview}%
  \BibitemOpen
  \bibfield  {author} {\bibinfo {author} {\bibfnamefont {L.}~\bibnamefont
  {Gammaitoni}}, \bibinfo {author} {\bibfnamefont {P.}~\bibnamefont
  {H\"anggi}}, \bibinfo {author} {\bibfnamefont {P.}~\bibnamefont {Jung}}, \
  and\ \bibinfo {author} {\bibfnamefont {F.}~\bibnamefont {Marchesoni}},\
  }\href {\doibase 10.1103/RevModPhys.70.223} {\bibfield  {journal} {\bibinfo
  {journal} {Rev. Mod. Phys.}\ }\textbf {\bibinfo {volume} {70}},\ \bibinfo
  {pages} {223} (\bibinfo {year} {1998})}\BibitemShut {NoStop}%
\bibitem [{\citenamefont {Chen}\ and\ \citenamefont {Xu}(2019)}]{chen2019prb}%
  \BibitemOpen
  \bibfield  {author} {\bibinfo {author} {\bibfnamefont {R.}~\bibnamefont
  {Chen}}\ and\ \bibinfo {author} {\bibfnamefont {X.}~\bibnamefont {Xu}},\
  }\href {\doibase 10.1103/PhysRevB.100.115437} {\bibfield  {journal} {\bibinfo
   {journal} {Phys. Rev. B}\ }\textbf {\bibinfo {volume} {100}},\ \bibinfo
  {pages} {115437} (\bibinfo {year} {2019})}\BibitemShut {NoStop}%
\bibitem [{\citenamefont {Kato}\ and\ \citenamefont
  {Nakao}(2021)}]{kato2021njp}%
  \BibitemOpen
  \bibfield  {author} {\bibinfo {author} {\bibfnamefont {Y.}~\bibnamefont
  {Kato}}\ and\ \bibinfo {author} {\bibfnamefont {H.}~\bibnamefont {Nakao}},\
  }\href {\doibase 10.1088/1367-2630/abf1d7} {\bibfield  {journal} {\bibinfo
  {journal} {New J. Phys.}\ }\textbf {\bibinfo {volume} {23}},\ \bibinfo
  {pages} {043018} (\bibinfo {year} {2021})}\BibitemShut {NoStop}%
\bibitem [{\citenamefont {Wagner}\ \emph {et~al.}(2019)\citenamefont {Wagner},
  \citenamefont {Talkner}, \citenamefont {Bayer}, \citenamefont {Rugeramigabo},
  \citenamefont {H\"anggi},\ and\ \citenamefont {Haug}}]{wagner2019nat}%
  \BibitemOpen
  \bibfield  {author} {\bibinfo {author} {\bibfnamefont {T.}~\bibnamefont
  {Wagner}}, \bibinfo {author} {\bibfnamefont {P.}~\bibnamefont {Talkner}},
  \bibinfo {author} {\bibfnamefont {J.~C.}\ \bibnamefont {Bayer}}, \bibinfo
  {author} {\bibfnamefont {E.~P.}\ \bibnamefont {Rugeramigabo}}, \bibinfo
  {author} {\bibfnamefont {P.}~\bibnamefont {H\"anggi}}, \ and\ \bibinfo
  {author} {\bibfnamefont {R.~J.}\ \bibnamefont {Haug}},\ }\href@noop {}
  {\bibfield  {journal} {\bibinfo  {journal} {Nat. Phys.}\ }\textbf {\bibinfo
  {volume} {15}},\ \bibinfo {pages} {330} (\bibinfo {year} {2019})}\BibitemShut
  {NoStop}%
\bibitem [{\citenamefont {Hussein}\ \emph {et~al.}(2020)\citenamefont
  {Hussein}, \citenamefont {Kohler}, \citenamefont {Bayer}, \citenamefont
  {Wagner},\ and\ \citenamefont {Haug}}]{hussein2020prl}%
  \BibitemOpen
  \bibfield  {author} {\bibinfo {author} {\bibfnamefont {R.}~\bibnamefont
  {Hussein}}, \bibinfo {author} {\bibfnamefont {S.}~\bibnamefont {Kohler}},
  \bibinfo {author} {\bibfnamefont {J.~C.}\ \bibnamefont {Bayer}}, \bibinfo
  {author} {\bibfnamefont {T.}~\bibnamefont {Wagner}}, \ and\ \bibinfo {author}
  {\bibfnamefont {R.~J.}\ \bibnamefont {Haug}},\ }\href {\doibase
  10.1103/PhysRevLett.125.206801} {\bibfield  {journal} {\bibinfo  {journal}
  {Phys. Rev. Lett.}\ }\textbf {\bibinfo {volume} {125}},\ \bibinfo {pages}
  {206801} (\bibinfo {year} {2020})}\BibitemShut {NoStop}%
\bibitem [{\citenamefont {Li}\ \emph {et~al.}(2018)\citenamefont {Li},
  \citenamefont {Chen},\ and\ \citenamefont {Fisher}}]{Li2018prb}%
  \BibitemOpen
  \bibfield  {author} {\bibinfo {author} {\bibfnamefont {Y.}~\bibnamefont
  {Li}}, \bibinfo {author} {\bibfnamefont {X.}~\bibnamefont {Chen}}, \ and\
  \bibinfo {author} {\bibfnamefont {M.~P.~A.}\ \bibnamefont {Fisher}},\ }\href
  {\doibase 10.1103/PhysRevB.98.205136} {\bibfield  {journal} {\bibinfo
  {journal} {Phys. Rev. B}\ }\textbf {\bibinfo {volume} {98}},\ \bibinfo
  {pages} {205136} (\bibinfo {year} {2018})}\BibitemShut {NoStop}%
\bibitem [{\citenamefont {Skinner}\ \emph {et~al.}(2019)\citenamefont
  {Skinner}, \citenamefont {Ruhman},\ and\ \citenamefont
  {Nahum}}]{skinner2019prx}%
  \BibitemOpen
  \bibfield  {author} {\bibinfo {author} {\bibfnamefont {B.}~\bibnamefont
  {Skinner}}, \bibinfo {author} {\bibfnamefont {J.}~\bibnamefont {Ruhman}}, \
  and\ \bibinfo {author} {\bibfnamefont {A.}~\bibnamefont {Nahum}},\ }\href
  {\doibase 10.1103/PhysRevX.9.031009} {\bibfield  {journal} {\bibinfo
  {journal} {Phys. Rev. X}\ }\textbf {\bibinfo {volume} {9}},\ \bibinfo {pages}
  {031009} (\bibinfo {year} {2019})}\BibitemShut {NoStop}%
\bibitem [{\citenamefont {Choi}\ \emph {et~al.}(2020)\citenamefont {Choi},
  \citenamefont {Bao}, \citenamefont {Qi},\ and\ \citenamefont
  {Altman}}]{Ehud1}%
  \BibitemOpen
  \bibfield  {author} {\bibinfo {author} {\bibfnamefont {S.}~\bibnamefont
  {Choi}}, \bibinfo {author} {\bibfnamefont {Y.}~\bibnamefont {Bao}}, \bibinfo
  {author} {\bibfnamefont {X.-L.}\ \bibnamefont {Qi}}, \ and\ \bibinfo {author}
  {\bibfnamefont {E.}~\bibnamefont {Altman}},\ }\href {\doibase
  10.1103/PhysRevLett.125.030505} {\bibfield  {journal} {\bibinfo  {journal}
  {Phys. Rev. Lett.}\ }\textbf {\bibinfo {volume} {125}},\ \bibinfo {pages}
  {030505} (\bibinfo {year} {2020})}\BibitemShut {NoStop}%
\bibitem [{\citenamefont {Bao}\ \emph {et~al.}(2021)\citenamefont {Bao},
  \citenamefont {Choi},\ and\ \citenamefont {Altman}}]{Ehud2}%
  \BibitemOpen
  \bibfield  {author} {\bibinfo {author} {\bibfnamefont {Y.}~\bibnamefont
  {Bao}}, \bibinfo {author} {\bibfnamefont {S.}~\bibnamefont {Choi}}, \ and\
  \bibinfo {author} {\bibfnamefont {E.}~\bibnamefont {Altman}},\ }\href@noop {}
  {\enquote {\bibinfo {title} {Symmetry enriched phases of quantum circuits},}\
  } (\bibinfo {year} {2021}),\ \Eprint {http://arxiv.org/abs/2102.09164}
  {arXiv:2102.09164 [cond-mat.stat-mech]} \BibitemShut {NoStop}%
\bibitem [{\citenamefont {Bloch}\ \emph {et~al.}(2008)\citenamefont {Bloch},
  \citenamefont {Dalibard},\ and\ \citenamefont {Zwerger}}]{coldatom1}%
  \BibitemOpen
  \bibfield  {author} {\bibinfo {author} {\bibfnamefont {I.}~\bibnamefont
  {Bloch}}, \bibinfo {author} {\bibfnamefont {J.}~\bibnamefont {Dalibard}}, \
  and\ \bibinfo {author} {\bibfnamefont {W.}~\bibnamefont {Zwerger}},\ }\href
  {\doibase 10.1103/RevModPhys.80.885} {\bibfield  {journal} {\bibinfo
  {journal} {Rev. Mod. Phys.}\ }\textbf {\bibinfo {volume} {80}},\ \bibinfo
  {pages} {885} (\bibinfo {year} {2008})}\BibitemShut {NoStop}%
\bibitem [{\citenamefont {Castro-Alvaredo}\ \emph {et~al.}(2020)\citenamefont
  {Castro-Alvaredo}, \citenamefont {Lencs\'es}, \citenamefont {Sz\'ecs\'enyi},\
  and\ \citenamefont {Viti}}]{Olalla}%
  \BibitemOpen
  \bibfield  {author} {\bibinfo {author} {\bibfnamefont {O.~A.}\ \bibnamefont
  {Castro-Alvaredo}}, \bibinfo {author} {\bibfnamefont {M.}~\bibnamefont
  {Lencs\'es}}, \bibinfo {author} {\bibfnamefont {I.~M.}\ \bibnamefont
  {Sz\'ecs\'enyi}}, \ and\ \bibinfo {author} {\bibfnamefont {J.}~\bibnamefont
  {Viti}},\ }\href {\doibase 10.1103/PhysRevLett.124.230601} {\bibfield
  {journal} {\bibinfo  {journal} {Phys. Rev. Lett.}\ }\textbf {\bibinfo
  {volume} {124}},\ \bibinfo {pages} {230601} (\bibinfo {year}
  {2020})}\BibitemShut {NoStop}%
\bibitem [{\citenamefont {Buča}\ \emph {et~al.}(2019)\citenamefont {Buča},
  \citenamefont {Tindall},\ and\ \citenamefont {Jaksch}}]{Buca_2019}%
  \BibitemOpen
  \bibfield  {author} {\bibinfo {author} {\bibfnamefont {B.}~\bibnamefont
  {Buča}}, \bibinfo {author} {\bibfnamefont {J.}~\bibnamefont {Tindall}}, \
  and\ \bibinfo {author} {\bibfnamefont {D.}~\bibnamefont {Jaksch}},\ }\href
  {\doibase 10.1038/s41467-019-09757-y} {\bibfield  {journal} {\bibinfo
  {journal} {Nature Communications}\ }\textbf {\bibinfo {volume} {10}},\
  \bibinfo {pages} {1730} (\bibinfo {year} {2019})}\BibitemShut {NoStop}%
\bibitem [{\citenamefont {Serbyn}\ \emph {et~al.}(2021)\citenamefont {Serbyn},
  \citenamefont {Abanin},\ and\ \citenamefont {Papić}}]{Serbyn_2021}%
  \BibitemOpen
  \bibfield  {author} {\bibinfo {author} {\bibfnamefont {M.}~\bibnamefont
  {Serbyn}}, \bibinfo {author} {\bibfnamefont {D.~A.}\ \bibnamefont {Abanin}},
  \ and\ \bibinfo {author} {\bibfnamefont {Z.}~\bibnamefont {Papić}},\ }\href
  {\doibase 10.1038/s41567-021-01230-2} {\bibfield  {journal} {\bibinfo
  {journal} {Nature Physics}\ } (\bibinfo {year} {2021}),\
  10.1038/s41567-021-01230-2}\BibitemShut {NoStop}%
\bibitem [{\citenamefont {Medenjak}\ \emph {et~al.}(2020)\citenamefont
  {Medenjak}, \citenamefont {Bu\ifmmode~\check{c}\else \v{c}\fi{}a},\ and\
  \citenamefont {Jaksch}}]{Marko1}%
  \BibitemOpen
  \bibfield  {author} {\bibinfo {author} {\bibfnamefont {M.}~\bibnamefont
  {Medenjak}}, \bibinfo {author} {\bibfnamefont {B.}~\bibnamefont
  {Bu\ifmmode~\check{c}\else \v{c}\fi{}a}}, \ and\ \bibinfo {author}
  {\bibfnamefont {D.}~\bibnamefont {Jaksch}},\ }\href {\doibase
  10.1103/PhysRevB.102.041117} {\bibfield  {journal} {\bibinfo  {journal}
  {Phys. Rev. B}\ }\textbf {\bibinfo {volume} {102}},\ \bibinfo {pages}
  {041117} (\bibinfo {year} {2020})}\BibitemShut {NoStop}%
\bibitem [{\citenamefont {Chinzei}\ and\ \citenamefont
  {Ikeda}(2020)}]{Chinzei}%
  \BibitemOpen
  \bibfield  {author} {\bibinfo {author} {\bibfnamefont {K.}~\bibnamefont
  {Chinzei}}\ and\ \bibinfo {author} {\bibfnamefont {T.~N.}\ \bibnamefont
  {Ikeda}},\ }\href {\doibase 10.1103/PhysRevLett.125.060601} {\bibfield
  {journal} {\bibinfo  {journal} {Phys. Rev. Lett.}\ }\textbf {\bibinfo
  {volume} {125}},\ \bibinfo {pages} {060601} (\bibinfo {year}
  {2020})}\BibitemShut {NoStop}%
\bibitem [{\citenamefont {Bu\ifmmode~\check{c}\else \v{c}\fi{}a}\ and\
  \citenamefont {Jaksch}(2019)}]{BucaJaksch2019}%
  \BibitemOpen
  \bibfield  {author} {\bibinfo {author} {\bibfnamefont {B.}~\bibnamefont
  {Bu\ifmmode~\check{c}\else \v{c}\fi{}a}}\ and\ \bibinfo {author}
  {\bibfnamefont {D.}~\bibnamefont {Jaksch}},\ }\href {\doibase
  10.1103/PhysRevLett.123.260401} {\bibfield  {journal} {\bibinfo  {journal}
  {Phys. Rev. Lett.}\ }\textbf {\bibinfo {volume} {123}},\ \bibinfo {pages}
  {260401} (\bibinfo {year} {2019})}\BibitemShut {NoStop}%
\bibitem [{\citenamefont {Booker}\ \emph {et~al.}(2020)\citenamefont {Booker},
  \citenamefont {Buča},\ and\ \citenamefont {Jaksch}}]{Booker_2020}%
  \BibitemOpen
  \bibfield  {author} {\bibinfo {author} {\bibfnamefont {C.}~\bibnamefont
  {Booker}}, \bibinfo {author} {\bibfnamefont {B.}~\bibnamefont {Buča}}, \
  and\ \bibinfo {author} {\bibfnamefont {D.}~\bibnamefont {Jaksch}},\ }\href
  {\doibase 10.1088/1367-2630/ababc4} {\bibfield  {journal} {\bibinfo
  {journal} {New Journal of Physics}\ } (\bibinfo {year} {2020}),\
  10.1088/1367-2630/ababc4}\BibitemShut {NoStop}%
\bibitem [{\citenamefont {S\'anchez Mu\~noz}\ \emph {et~al.}(2019)\citenamefont
  {S\'anchez Mu\~noz}, \citenamefont {Bu\ifmmode~\check{c}\else \v{c}\fi{}a},
  \citenamefont {Tindall}, \citenamefont {Gonz\'alez-Tudela}, \citenamefont
  {Jaksch},\ and\ \citenamefont {Porras}}]{Carlos1}%
  \BibitemOpen
  \bibfield  {author} {\bibinfo {author} {\bibfnamefont {C.}~\bibnamefont
  {S\'anchez Mu\~noz}}, \bibinfo {author} {\bibfnamefont {B.}~\bibnamefont
  {Bu\ifmmode~\check{c}\else \v{c}\fi{}a}}, \bibinfo {author} {\bibfnamefont
  {J.}~\bibnamefont {Tindall}}, \bibinfo {author} {\bibfnamefont
  {A.}~\bibnamefont {Gonz\'alez-Tudela}}, \bibinfo {author} {\bibfnamefont
  {D.}~\bibnamefont {Jaksch}}, \ and\ \bibinfo {author} {\bibfnamefont
  {D.}~\bibnamefont {Porras}},\ }\href {\doibase 10.1103/PhysRevA.100.042113}
  {\bibfield  {journal} {\bibinfo  {journal} {Phys. Rev. A}\ }\textbf {\bibinfo
  {volume} {100}},\ \bibinfo {pages} {042113} (\bibinfo {year}
  {2019})}\BibitemShut {NoStop}%
\bibitem [{\citenamefont {Tucker}\ \emph {et~al.}(2018)\citenamefont {Tucker},
  \citenamefont {Zhu}, \citenamefont {Lewis-Swan}, \citenamefont {Marino},
  \citenamefont {Jimenez}, \citenamefont {Restrepo},\ and\ \citenamefont
  {Rey}}]{Jamir1}%
  \BibitemOpen
  \bibfield  {author} {\bibinfo {author} {\bibfnamefont {K.}~\bibnamefont
  {Tucker}}, \bibinfo {author} {\bibfnamefont {B.}~\bibnamefont {Zhu}},
  \bibinfo {author} {\bibfnamefont {R.~J.}\ \bibnamefont {Lewis-Swan}},
  \bibinfo {author} {\bibfnamefont {J.}~\bibnamefont {Marino}}, \bibinfo
  {author} {\bibfnamefont {F.}~\bibnamefont {Jimenez}}, \bibinfo {author}
  {\bibfnamefont {J.~G.}\ \bibnamefont {Restrepo}}, \ and\ \bibinfo {author}
  {\bibfnamefont {A.~M.}\ \bibnamefont {Rey}},\ }\href@noop {} {\bibfield
  {journal} {\bibinfo  {journal} {New Journal of Physics}\ }\textbf {\bibinfo
  {volume} {20}},\ \bibinfo {pages} {123003} (\bibinfo {year}
  {2018})}\BibitemShut {NoStop}%
\bibitem [{\citenamefont {Ke{\ss}ler}\ \emph {et~al.}(2020)\citenamefont
  {Ke{\ss}ler}, \citenamefont {Kongkhambut}, \citenamefont {Georges},
  \citenamefont {Mathey}, \citenamefont {Cosme},\ and\ \citenamefont
  {Hemmerich}}]{dissipativeTCobs}%
  \BibitemOpen
  \bibfield  {author} {\bibinfo {author} {\bibfnamefont {H.}~\bibnamefont
  {Ke{\ss}ler}}, \bibinfo {author} {\bibfnamefont {P.}~\bibnamefont
  {Kongkhambut}}, \bibinfo {author} {\bibfnamefont {C.}~\bibnamefont
  {Georges}}, \bibinfo {author} {\bibfnamefont {L.}~\bibnamefont {Mathey}},
  \bibinfo {author} {\bibfnamefont {J.~G.}\ \bibnamefont {Cosme}}, \ and\
  \bibinfo {author} {\bibfnamefont {A.}~\bibnamefont {Hemmerich}},\ }\href@noop
  {} {\bibfield  {journal} {\bibinfo  {journal} {arXiv preprint
  arXiv:2012.08885}\ } (\bibinfo {year} {2020})}\BibitemShut {NoStop}%
\bibitem [{\citenamefont {Dogra}\ \emph {et~al.}(2019)\citenamefont {Dogra},
  \citenamefont {Landini}, \citenamefont {Kroeger}, \citenamefont {Hruby},
  \citenamefont {Donner},\ and\ \citenamefont {Esslinger}}]{Esslinger}%
  \BibitemOpen
  \bibfield  {author} {\bibinfo {author} {\bibfnamefont {N.}~\bibnamefont
  {Dogra}}, \bibinfo {author} {\bibfnamefont {M.}~\bibnamefont {Landini}},
  \bibinfo {author} {\bibfnamefont {K.}~\bibnamefont {Kroeger}}, \bibinfo
  {author} {\bibfnamefont {L.}~\bibnamefont {Hruby}}, \bibinfo {author}
  {\bibfnamefont {T.}~\bibnamefont {Donner}}, \ and\ \bibinfo {author}
  {\bibfnamefont {T.}~\bibnamefont {Esslinger}},\ }\href {\doibase
  10.1126/science.aaw4465} {\bibfield  {journal} {\bibinfo  {journal}
  {Science}\ }\textbf {\bibinfo {volume} {366}},\ \bibinfo {pages} {1496}
  (\bibinfo {year} {2019})},\ \Eprint
  {http://arxiv.org/abs/https://science.sciencemag.org/content/366/6472/1496.full.pdf}
  {https://science.sciencemag.org/content/366/6472/1496.full.pdf} \BibitemShut
  {NoStop}%
\bibitem [{\citenamefont {Bull}\ \emph {et~al.}(2020)\citenamefont {Bull},
  \citenamefont {Desaules},\ and\ \citenamefont {Papi\ifmmode~\acute{c}\else
  \'{c}\fi{}}}]{scarsdynsym1}%
  \BibitemOpen
  \bibfield  {author} {\bibinfo {author} {\bibfnamefont {K.}~\bibnamefont
  {Bull}}, \bibinfo {author} {\bibfnamefont {J.-Y.}\ \bibnamefont {Desaules}},
  \ and\ \bibinfo {author} {\bibfnamefont {Z.}~\bibnamefont
  {Papi\ifmmode~\acute{c}\else \'{c}\fi{}}},\ }\href {\doibase
  10.1103/PhysRevB.101.165139} {\bibfield  {journal} {\bibinfo  {journal}
  {Phys. Rev. B}\ }\textbf {\bibinfo {volume} {101}},\ \bibinfo {pages}
  {165139} (\bibinfo {year} {2020})}\BibitemShut {NoStop}%
\bibitem [{\citenamefont {Moudgalya}\ \emph {et~al.}(2020)\citenamefont
  {Moudgalya}, \citenamefont {Regnault},\ and\ \citenamefont
  {Bernevig}}]{scarsdynsym2}%
  \BibitemOpen
  \bibfield  {author} {\bibinfo {author} {\bibfnamefont {S.}~\bibnamefont
  {Moudgalya}}, \bibinfo {author} {\bibfnamefont {N.}~\bibnamefont {Regnault}},
  \ and\ \bibinfo {author} {\bibfnamefont {B.~A.}\ \bibnamefont {Bernevig}},\
  }\href {\doibase 10.1103/PhysRevB.102.085140} {\bibfield  {journal} {\bibinfo
   {journal} {Phys. Rev. B}\ }\textbf {\bibinfo {volume} {102}},\ \bibinfo
  {pages} {085140} (\bibinfo {year} {2020})}\BibitemShut {NoStop}%
\bibitem [{\citenamefont {Mark}\ and\ \citenamefont
  {Motrunich}(2020)}]{scarsdynsym3}%
  \BibitemOpen
  \bibfield  {author} {\bibinfo {author} {\bibfnamefont {D.~K.}\ \bibnamefont
  {Mark}}\ and\ \bibinfo {author} {\bibfnamefont {O.~I.}\ \bibnamefont
  {Motrunich}},\ }\href {\doibase 10.1103/PhysRevB.102.075132} {\bibfield
  {journal} {\bibinfo  {journal} {Phys. Rev. B}\ }\textbf {\bibinfo {volume}
  {102}},\ \bibinfo {pages} {075132} (\bibinfo {year} {2020})}\BibitemShut
  {NoStop}%
\bibitem [{\citenamefont {Mark}\ \emph {et~al.}(2020)\citenamefont {Mark},
  \citenamefont {Lin},\ and\ \citenamefont {Motrunich}}]{scarsdynsym4}%
  \BibitemOpen
  \bibfield  {author} {\bibinfo {author} {\bibfnamefont {D.~K.}\ \bibnamefont
  {Mark}}, \bibinfo {author} {\bibfnamefont {C.-J.}\ \bibnamefont {Lin}}, \
  and\ \bibinfo {author} {\bibfnamefont {O.~I.}\ \bibnamefont {Motrunich}},\
  }\href {http://dx.doi.org/10.1103/PhysRevB.101.195131} {\bibfield  {journal}
  {\bibinfo  {journal} {Phys. Rev. B}\ }\textbf {\bibinfo {volume} {101}}
  (\bibinfo {year} {2020})}\BibitemShut {NoStop}%
\bibitem [{\citenamefont {O'Dea}\ \emph {et~al.}(2020)\citenamefont {O'Dea},
  \citenamefont {Burnell}, \citenamefont {Chandran},\ and\ \citenamefont
  {Khemani}}]{scarsdynsym5}%
  \BibitemOpen
  \bibfield  {author} {\bibinfo {author} {\bibfnamefont {N.}~\bibnamefont
  {O'Dea}}, \bibinfo {author} {\bibfnamefont {F.}~\bibnamefont {Burnell}},
  \bibinfo {author} {\bibfnamefont {A.}~\bibnamefont {Chandran}}, \ and\
  \bibinfo {author} {\bibfnamefont {V.}~\bibnamefont {Khemani}},\ }\href
  {\doibase 10.1103/PhysRevResearch.2.043305} {\bibfield  {journal} {\bibinfo
  {journal} {Phys. Rev. Research}\ }\textbf {\bibinfo {volume} {2}},\ \bibinfo
  {pages} {043305} (\bibinfo {year} {2020})}\BibitemShut {NoStop}%
\bibitem [{\citenamefont {Pakrouski}\ \emph {et~al.}(2020)\citenamefont
  {Pakrouski}, \citenamefont {Pallegar}, \citenamefont {Popov},\ and\
  \citenamefont {Klebanov}}]{scarsdynsym6}%
  \BibitemOpen
  \bibfield  {author} {\bibinfo {author} {\bibfnamefont {K.}~\bibnamefont
  {Pakrouski}}, \bibinfo {author} {\bibfnamefont {P.~N.}\ \bibnamefont
  {Pallegar}}, \bibinfo {author} {\bibfnamefont {F.~K.}\ \bibnamefont {Popov}},
  \ and\ \bibinfo {author} {\bibfnamefont {I.~R.}\ \bibnamefont {Klebanov}},\
  }\href {\doibase 10.1103/PhysRevLett.125.230602} {\bibfield  {journal}
  {\bibinfo  {journal} {Phys. Rev. Lett.}\ }\textbf {\bibinfo {volume} {125}},\
  \bibinfo {pages} {230602} (\bibinfo {year} {2020})}\BibitemShut {NoStop}%
\bibitem [{\citenamefont {Buca}\ \emph {et~al.}(2020)\citenamefont {Buca},
  \citenamefont {Purkayastha}, \citenamefont {Guarnieri}, \citenamefont
  {Mitchison}, \citenamefont {Jaksch},\ and\ \citenamefont
  {Goold}}]{Buca_2020}%
  \BibitemOpen
  \bibfield  {author} {\bibinfo {author} {\bibfnamefont {B.}~\bibnamefont
  {Buca}}, \bibinfo {author} {\bibfnamefont {A.}~\bibnamefont {Purkayastha}},
  \bibinfo {author} {\bibfnamefont {G.}~\bibnamefont {Guarnieri}}, \bibinfo
  {author} {\bibfnamefont {M.~T.}\ \bibnamefont {Mitchison}}, \bibinfo {author}
  {\bibfnamefont {D.}~\bibnamefont {Jaksch}}, \ and\ \bibinfo {author}
  {\bibfnamefont {J.}~\bibnamefont {Goold}},\ }\href@noop {} {\enquote
  {\bibinfo {title} {Quantum many-body attractors},}\ } (\bibinfo {year}
  {2020}),\ \Eprint {http://arxiv.org/abs/2008.11166} {arXiv:2008.11166
  [quant-ph]} \BibitemShut {NoStop}%
\bibitem [{\citenamefont {Tindall}\ \emph {et~al.}(2020)\citenamefont
  {Tindall}, \citenamefont {Sánchez~Muñoz}, \citenamefont {Buča},\ and\
  \citenamefont {Jaksch}}]{quantumsynch}%
  \BibitemOpen
  \bibfield  {author} {\bibinfo {author} {\bibfnamefont {J.}~\bibnamefont
  {Tindall}}, \bibinfo {author} {\bibfnamefont {C.}~\bibnamefont
  {Sánchez~Muñoz}}, \bibinfo {author} {\bibfnamefont {B.}~\bibnamefont
  {Buča}}, \ and\ \bibinfo {author} {\bibfnamefont {D.}~\bibnamefont
  {Jaksch}},\ }\href {\doibase 10.1088/1367-2630/ab60f5} {\bibfield  {journal}
  {\bibinfo  {journal} {New Journal of Physics}\ }\textbf {\bibinfo {volume}
  {22}},\ \bibinfo {pages} {013026} (\bibinfo {year} {2020})}\BibitemShut
  {NoStop}%
\bibitem [{\citenamefont {Buca}\ \emph {et~al.}(2021)\citenamefont {Buca},
  \citenamefont {Booker},\ and\ \citenamefont {Jaksch}}]{buca2021algebraic}%
  \BibitemOpen
  \bibfield  {author} {\bibinfo {author} {\bibfnamefont {B.}~\bibnamefont
  {Buca}}, \bibinfo {author} {\bibfnamefont {C.}~\bibnamefont {Booker}}, \ and\
  \bibinfo {author} {\bibfnamefont {D.}~\bibnamefont {Jaksch}},\ }\href@noop {}
  {\enquote {\bibinfo {title} {Algebraic theory of quantum synchronization and
  limit cycles under dissipation},}\ } (\bibinfo {year} {2021}),\ \Eprint
  {http://arxiv.org/abs/2103.01808} {arXiv:2103.01808 [quant-ph]} \BibitemShut
  {NoStop}%
\bibitem [{\citenamefont {Tindall}\ \emph {et~al.}(2019)\citenamefont
  {Tindall}, \citenamefont {Bu\ifmmode~\check{c}\else \v{c}\fi{}a},
  \citenamefont {Coulthard},\ and\ \citenamefont
  {Jaksch}}]{PhysRevLett.123.030603}%
  \BibitemOpen
  \bibfield  {author} {\bibinfo {author} {\bibfnamefont {J.}~\bibnamefont
  {Tindall}}, \bibinfo {author} {\bibfnamefont {B.}~\bibnamefont
  {Bu\ifmmode~\check{c}\else \v{c}\fi{}a}}, \bibinfo {author} {\bibfnamefont
  {J.~R.}\ \bibnamefont {Coulthard}}, \ and\ \bibinfo {author} {\bibfnamefont
  {D.}~\bibnamefont {Jaksch}},\ }\href {\doibase
  10.1103/PhysRevLett.123.030603} {\bibfield  {journal} {\bibinfo  {journal}
  {Phys. Rev. Lett.}\ }\textbf {\bibinfo {volume} {123}},\ \bibinfo {pages}
  {030603} (\bibinfo {year} {2019})}\BibitemShut {NoStop}%
\bibitem [{\citenamefont {Sugawa}\ \emph {et~al.}(2011)\citenamefont {Sugawa},
  \citenamefont {Inaba}, \citenamefont {Taie}, \citenamefont {Yamazaki},
  \citenamefont {Yamashita},\ and\ \citenamefont {Takahashi}}]{sugawa2011nat}%
  \BibitemOpen
  \bibfield  {author} {\bibinfo {author} {\bibfnamefont {S.}~\bibnamefont
  {Sugawa}}, \bibinfo {author} {\bibfnamefont {K.}~\bibnamefont {Inaba}},
  \bibinfo {author} {\bibfnamefont {S.}~\bibnamefont {Taie}}, \bibinfo {author}
  {\bibfnamefont {R.}~\bibnamefont {Yamazaki}}, \bibinfo {author}
  {\bibfnamefont {M.}~\bibnamefont {Yamashita}}, \ and\ \bibinfo {author}
  {\bibfnamefont {Y.}~\bibnamefont {Takahashi}},\ }\href@noop {} {\bibfield
  {journal} {\bibinfo  {journal} {Nat. Phys.}\ }\textbf {\bibinfo {volume}
  {7}},\ \bibinfo {pages} {642} (\bibinfo {year} {2011})}\BibitemShut {NoStop}%
\bibitem [{\citenamefont {Ferrier-Barbut}\ \emph {et~al.}(2014)\citenamefont
  {Ferrier-Barbut}, \citenamefont {Delehaye}, \citenamefont {Laurent},
  \citenamefont {Grier}, \citenamefont {Pierce}, \citenamefont {Rem},
  \citenamefont {Chevy},\ and\ \citenamefont {Salomon}}]{ferrier2014science}%
  \BibitemOpen
  \bibfield  {author} {\bibinfo {author} {\bibfnamefont {I.}~\bibnamefont
  {Ferrier-Barbut}}, \bibinfo {author} {\bibfnamefont {M.}~\bibnamefont
  {Delehaye}}, \bibinfo {author} {\bibfnamefont {S.}~\bibnamefont {Laurent}},
  \bibinfo {author} {\bibfnamefont {A.~T.}\ \bibnamefont {Grier}}, \bibinfo
  {author} {\bibfnamefont {M.}~\bibnamefont {Pierce}}, \bibinfo {author}
  {\bibfnamefont {B.~S.}\ \bibnamefont {Rem}}, \bibinfo {author} {\bibfnamefont
  {F.}~\bibnamefont {Chevy}}, \ and\ \bibinfo {author} {\bibfnamefont
  {C.}~\bibnamefont {Salomon}},\ }\href@noop {} {\bibfield  {journal} {\bibinfo
   {journal} {Science}\ }\textbf {\bibinfo {volume} {345}},\ \bibinfo {pages}
  {1035} (\bibinfo {year} {2014})}\BibitemShut {NoStop}%
\bibitem [{\citenamefont {Delehaye}\ \emph {et~al.}(2015)\citenamefont
  {Delehaye}, \citenamefont {Laurent}, \citenamefont {Ferrier-Barbut},
  \citenamefont {Jin}, \citenamefont {Chevy},\ and\ \citenamefont
  {Salomon}}]{delehave2015prl}%
  \BibitemOpen
  \bibfield  {author} {\bibinfo {author} {\bibfnamefont {M.}~\bibnamefont
  {Delehaye}}, \bibinfo {author} {\bibfnamefont {S.}~\bibnamefont {Laurent}},
  \bibinfo {author} {\bibfnamefont {I.}~\bibnamefont {Ferrier-Barbut}},
  \bibinfo {author} {\bibfnamefont {S.}~\bibnamefont {Jin}}, \bibinfo {author}
  {\bibfnamefont {F.}~\bibnamefont {Chevy}}, \ and\ \bibinfo {author}
  {\bibfnamefont {C.}~\bibnamefont {Salomon}},\ }\href {\doibase
  10.1103/PhysRevLett.115.265303} {\bibfield  {journal} {\bibinfo  {journal}
  {Phys. Rev. Lett.}\ }\textbf {\bibinfo {volume} {115}},\ \bibinfo {pages}
  {265303} (\bibinfo {year} {2015})}\BibitemShut {NoStop}%
\bibitem [{\citenamefont {Roy}\ \emph {et~al.}(2017)\citenamefont {Roy},
  \citenamefont {Green}, \citenamefont {Bowler},\ and\ \citenamefont
  {Gupta}}]{roy2017prl}%
  \BibitemOpen
  \bibfield  {author} {\bibinfo {author} {\bibfnamefont {R.}~\bibnamefont
  {Roy}}, \bibinfo {author} {\bibfnamefont {A.}~\bibnamefont {Green}}, \bibinfo
  {author} {\bibfnamefont {R.}~\bibnamefont {Bowler}}, \ and\ \bibinfo {author}
  {\bibfnamefont {S.}~\bibnamefont {Gupta}},\ }\href {\doibase
  10.1103/PhysRevLett.118.055301} {\bibfield  {journal} {\bibinfo  {journal}
  {Phys. Rev. Lett.}\ }\textbf {\bibinfo {volume} {118}},\ \bibinfo {pages}
  {055301} (\bibinfo {year} {2017})}\BibitemShut {NoStop}%
\bibitem [{\citenamefont {Lous}\ \emph {et~al.}(2018)\citenamefont {Lous},
  \citenamefont {Fritsche}, \citenamefont {Jag}, \citenamefont {Lehmann},
  \citenamefont {Kirilov}, \citenamefont {Huang},\ and\ \citenamefont
  {Grimm}}]{lous2018prl}%
  \BibitemOpen
  \bibfield  {author} {\bibinfo {author} {\bibfnamefont {R.~S.}\ \bibnamefont
  {Lous}}, \bibinfo {author} {\bibfnamefont {I.}~\bibnamefont {Fritsche}},
  \bibinfo {author} {\bibfnamefont {M.}~\bibnamefont {Jag}}, \bibinfo {author}
  {\bibfnamefont {F.}~\bibnamefont {Lehmann}}, \bibinfo {author} {\bibfnamefont
  {E.}~\bibnamefont {Kirilov}}, \bibinfo {author} {\bibfnamefont
  {B.}~\bibnamefont {Huang}}, \ and\ \bibinfo {author} {\bibfnamefont
  {R.}~\bibnamefont {Grimm}},\ }\href {\doibase 10.1103/PhysRevLett.120.243403}
  {\bibfield  {journal} {\bibinfo  {journal} {Phys. Rev. Lett.}\ }\textbf
  {\bibinfo {volume} {120}},\ \bibinfo {pages} {243403} (\bibinfo {year}
  {2018})}\BibitemShut {NoStop}%
\bibitem [{\citenamefont {Trautmann}\ \emph {et~al.}(2018)\citenamefont
  {Trautmann}, \citenamefont {Ilzh\"ofer}, \citenamefont {Durastante},
  \citenamefont {Politi}, \citenamefont {Sohmen}, \citenamefont {Mark},\ and\
  \citenamefont {Ferlaino}}]{trautmann2018prl}%
  \BibitemOpen
  \bibfield  {author} {\bibinfo {author} {\bibfnamefont {A.}~\bibnamefont
  {Trautmann}}, \bibinfo {author} {\bibfnamefont {P.}~\bibnamefont
  {Ilzh\"ofer}}, \bibinfo {author} {\bibfnamefont {G.}~\bibnamefont
  {Durastante}}, \bibinfo {author} {\bibfnamefont {C.}~\bibnamefont {Politi}},
  \bibinfo {author} {\bibfnamefont {M.}~\bibnamefont {Sohmen}}, \bibinfo
  {author} {\bibfnamefont {M.~J.}\ \bibnamefont {Mark}}, \ and\ \bibinfo
  {author} {\bibfnamefont {F.}~\bibnamefont {Ferlaino}},\ }\href {\doibase
  10.1103/PhysRevLett.121.213601} {\bibfield  {journal} {\bibinfo  {journal}
  {Phys. Rev. Lett.}\ }\textbf {\bibinfo {volume} {121}},\ \bibinfo {pages}
  {213601} (\bibinfo {year} {2018})}\BibitemShut {NoStop}%
\bibitem [{\citenamefont {Albus}\ \emph {et~al.}(2003)\citenamefont {Albus},
  \citenamefont {Illuminati},\ and\ \citenamefont {Eisert}}]{albus2003pra}%
  \BibitemOpen
  \bibfield  {author} {\bibinfo {author} {\bibfnamefont {A.}~\bibnamefont
  {Albus}}, \bibinfo {author} {\bibfnamefont {F.}~\bibnamefont {Illuminati}}, \
  and\ \bibinfo {author} {\bibfnamefont {J.}~\bibnamefont {Eisert}},\ }\href
  {\doibase 10.1103/PhysRevA.68.023606} {\bibfield  {journal} {\bibinfo
  {journal} {Phys. Rev. A}\ }\textbf {\bibinfo {volume} {68}},\ \bibinfo
  {pages} {023606} (\bibinfo {year} {2003})}\BibitemShut {NoStop}%
\bibitem [{\citenamefont {Lewenstein}\ \emph {et~al.}(2004)\citenamefont
  {Lewenstein}, \citenamefont {Santos}, \citenamefont {Baranov},\ and\
  \citenamefont {Fehrmann}}]{lewenstein2004prl}%
  \BibitemOpen
  \bibfield  {author} {\bibinfo {author} {\bibfnamefont {M.}~\bibnamefont
  {Lewenstein}}, \bibinfo {author} {\bibfnamefont {L.}~\bibnamefont {Santos}},
  \bibinfo {author} {\bibfnamefont {M.~A.}\ \bibnamefont {Baranov}}, \ and\
  \bibinfo {author} {\bibfnamefont {H.}~\bibnamefont {Fehrmann}},\ }\href
  {\doibase 10.1103/PhysRevLett.92.050401} {\bibfield  {journal} {\bibinfo
  {journal} {Phys. Rev. Lett.}\ }\textbf {\bibinfo {volume} {92}},\ \bibinfo
  {pages} {050401} (\bibinfo {year} {2004})}\BibitemShut {NoStop}%
\bibitem [{\citenamefont {Pollet}\ \emph {et~al.}(2006)\citenamefont {Pollet},
  \citenamefont {Troyer}, \citenamefont {Van~Houcke},\ and\ \citenamefont
  {Rombouts}}]{pollet2006prl}%
  \BibitemOpen
  \bibfield  {author} {\bibinfo {author} {\bibfnamefont {L.}~\bibnamefont
  {Pollet}}, \bibinfo {author} {\bibfnamefont {M.}~\bibnamefont {Troyer}},
  \bibinfo {author} {\bibfnamefont {K.}~\bibnamefont {Van~Houcke}}, \ and\
  \bibinfo {author} {\bibfnamefont {S.~M.~A.}\ \bibnamefont {Rombouts}},\
  }\href {\doibase 10.1103/PhysRevLett.96.190402} {\bibfield  {journal}
  {\bibinfo  {journal} {Phys. Rev. Lett.}\ }\textbf {\bibinfo {volume} {96}},\
  \bibinfo {pages} {190402} (\bibinfo {year} {2006})}\BibitemShut {NoStop}%
\bibitem [{\citenamefont {Zujev}\ \emph {et~al.}(2008)\citenamefont {Zujev},
  \citenamefont {Baldwin}, \citenamefont {Scalettar}, \citenamefont {Rousseau},
  \citenamefont {Denteneer},\ and\ \citenamefont {Rigol}}]{zujev2008pra}%
  \BibitemOpen
  \bibfield  {author} {\bibinfo {author} {\bibfnamefont {A.}~\bibnamefont
  {Zujev}}, \bibinfo {author} {\bibfnamefont {A.}~\bibnamefont {Baldwin}},
  \bibinfo {author} {\bibfnamefont {R.~T.}\ \bibnamefont {Scalettar}}, \bibinfo
  {author} {\bibfnamefont {V.~G.}\ \bibnamefont {Rousseau}}, \bibinfo {author}
  {\bibfnamefont {P.~J.~H.}\ \bibnamefont {Denteneer}}, \ and\ \bibinfo
  {author} {\bibfnamefont {M.}~\bibnamefont {Rigol}},\ }\href {\doibase
  10.1103/PhysRevA.78.033619} {\bibfield  {journal} {\bibinfo  {journal} {Phys.
  Rev. A}\ }\textbf {\bibinfo {volume} {78}},\ \bibinfo {pages} {033619}
  (\bibinfo {year} {2008})}\BibitemShut {NoStop}%
\bibitem [{\citenamefont {L\"uhmann}\ \emph {et~al.}(2008)\citenamefont
  {L\"uhmann}, \citenamefont {Bongs}, \citenamefont {Sengstock},\ and\
  \citenamefont {Pfannkuche}}]{luhmann2008prl}%
  \BibitemOpen
  \bibfield  {author} {\bibinfo {author} {\bibfnamefont {D.-S.}\ \bibnamefont
  {L\"uhmann}}, \bibinfo {author} {\bibfnamefont {K.}~\bibnamefont {Bongs}},
  \bibinfo {author} {\bibfnamefont {K.}~\bibnamefont {Sengstock}}, \ and\
  \bibinfo {author} {\bibfnamefont {D.}~\bibnamefont {Pfannkuche}},\ }\href
  {\doibase 10.1103/PhysRevLett.101.050402} {\bibfield  {journal} {\bibinfo
  {journal} {Phys. Rev. Lett.}\ }\textbf {\bibinfo {volume} {101}},\ \bibinfo
  {pages} {050402} (\bibinfo {year} {2008})}\BibitemShut {NoStop}%
\bibitem [{\citenamefont {Anders}\ \emph {et~al.}(2012)\citenamefont {Anders},
  \citenamefont {Werner}, \citenamefont {Troyer}, \citenamefont {Sigrist},\
  and\ \citenamefont {Pollet}}]{anders2012prl}%
  \BibitemOpen
  \bibfield  {author} {\bibinfo {author} {\bibfnamefont {P.}~\bibnamefont
  {Anders}}, \bibinfo {author} {\bibfnamefont {P.}~\bibnamefont {Werner}},
  \bibinfo {author} {\bibfnamefont {M.}~\bibnamefont {Troyer}}, \bibinfo
  {author} {\bibfnamefont {M.}~\bibnamefont {Sigrist}}, \ and\ \bibinfo
  {author} {\bibfnamefont {L.}~\bibnamefont {Pollet}},\ }\href {\doibase
  10.1103/PhysRevLett.109.206401} {\bibfield  {journal} {\bibinfo  {journal}
  {Phys. Rev. Lett.}\ }\textbf {\bibinfo {volume} {109}},\ \bibinfo {pages}
  {206401} (\bibinfo {year} {2012})}\BibitemShut {NoStop}%
\bibitem [{\citenamefont {Bukov}\ and\ \citenamefont
  {Pollet}(2014)}]{bukov2014prb}%
  \BibitemOpen
  \bibfield  {author} {\bibinfo {author} {\bibfnamefont {M.}~\bibnamefont
  {Bukov}}\ and\ \bibinfo {author} {\bibfnamefont {L.}~\bibnamefont {Pollet}},\
  }\href {\doibase 10.1103/PhysRevB.89.094502} {\bibfield  {journal} {\bibinfo
  {journal} {Phys. Rev. B}\ }\textbf {\bibinfo {volume} {89}},\ \bibinfo
  {pages} {094502} (\bibinfo {year} {2014})}\BibitemShut {NoStop}%
\bibitem [{\citenamefont {Avella}\ \emph {et~al.}(2019)\citenamefont {Avella},
  \citenamefont {Mendoza-Arenas}, \citenamefont {Franco},\ and\ \citenamefont
  {Silva-Valencia}}]{jj2019pra}%
  \BibitemOpen
  \bibfield  {author} {\bibinfo {author} {\bibfnamefont {R.}~\bibnamefont
  {Avella}}, \bibinfo {author} {\bibfnamefont {J.~J.}\ \bibnamefont
  {Mendoza-Arenas}}, \bibinfo {author} {\bibfnamefont {R.}~\bibnamefont
  {Franco}}, \ and\ \bibinfo {author} {\bibfnamefont {J.}~\bibnamefont
  {Silva-Valencia}},\ }\href {\doibase 10.1103/PhysRevA.100.063620} {\bibfield
  {journal} {\bibinfo  {journal} {Phys. Rev. A}\ }\textbf {\bibinfo {volume}
  {100}},\ \bibinfo {pages} {063620} (\bibinfo {year} {2019})}\BibitemShut
  {NoStop}%
\bibitem [{\citenamefont {Avella}\ \emph {et~al.}(2020)\citenamefont {Avella},
  \citenamefont {Mendoza-Arenas}, \citenamefont {Franco},\ and\ \citenamefont
  {Silva-Valencia}}]{jj2020pra}%
  \BibitemOpen
  \bibfield  {author} {\bibinfo {author} {\bibfnamefont {R.}~\bibnamefont
  {Avella}}, \bibinfo {author} {\bibfnamefont {J.~J.}\ \bibnamefont
  {Mendoza-Arenas}}, \bibinfo {author} {\bibfnamefont {R.}~\bibnamefont
  {Franco}}, \ and\ \bibinfo {author} {\bibfnamefont {J.}~\bibnamefont
  {Silva-Valencia}},\ }\href {\doibase 10.1103/PhysRevA.102.033341} {\bibfield
  {journal} {\bibinfo  {journal} {Phys. Rev. A}\ }\textbf {\bibinfo {volume}
  {102}},\ \bibinfo {pages} {033341} (\bibinfo {year} {2020})}\BibitemShut
  {NoStop}%
\bibitem [{\citenamefont {Guerrero-Suarez}\ \emph {et~al.}(2021)\citenamefont
  {Guerrero-Suarez}, \citenamefont {Mendoza-Arenas}, \citenamefont {Franco},\
  and\ \citenamefont {Silva-Valencia}}]{jj2021pra}%
  \BibitemOpen
  \bibfield  {author} {\bibinfo {author} {\bibfnamefont {R.}~\bibnamefont
  {Guerrero-Suarez}}, \bibinfo {author} {\bibfnamefont {J.~J.}\ \bibnamefont
  {Mendoza-Arenas}}, \bibinfo {author} {\bibfnamefont {R.}~\bibnamefont
  {Franco}}, \ and\ \bibinfo {author} {\bibfnamefont {J.}~\bibnamefont
  {Silva-Valencia}},\ }\href {\doibase 10.1103/PhysRevA.103.023304} {\bibfield
  {journal} {\bibinfo  {journal} {Phys. Rev. A}\ }\textbf {\bibinfo {volume}
  {103}},\ \bibinfo {pages} {023304} (\bibinfo {year} {2021})}\BibitemShut
  {NoStop}%
\bibitem [{\citenamefont {Muñoz}\ \emph {et~al.}(2019)\citenamefont {Muñoz},
  \citenamefont {Buča}, \citenamefont {Tindall}, \citenamefont
  {González-Tudela}, \citenamefont {Jaksch},\ and\ \citenamefont
  {Porras}}]{Carlos2}%
  \BibitemOpen
  \bibfield  {author} {\bibinfo {author} {\bibfnamefont {C.~S.}\ \bibnamefont
  {Muñoz}}, \bibinfo {author} {\bibfnamefont {B.}~\bibnamefont {Buča}},
  \bibinfo {author} {\bibfnamefont {J.}~\bibnamefont {Tindall}}, \bibinfo
  {author} {\bibfnamefont {A.}~\bibnamefont {González-Tudela}}, \bibinfo
  {author} {\bibfnamefont {D.}~\bibnamefont {Jaksch}}, \ and\ \bibinfo {author}
  {\bibfnamefont {D.}~\bibnamefont {Porras}},\ }\href@noop {} {\enquote
  {\bibinfo {title} {Non-stationary dynamics and dissipative freezing in
  squeezed superradiance},}\ } (\bibinfo {year} {2019}),\ \Eprint
  {http://arxiv.org/abs/1903.05080} {arXiv:1903.05080 [quant-ph]} \BibitemShut
  {NoStop}%
\bibitem [{\citenamefont {Halati}\ \emph {et~al.}(2021)\citenamefont {Halati},
  \citenamefont {Sheikhan},\ and\ \citenamefont {Kollath}}]{Kollath}%
  \BibitemOpen
  \bibfield  {author} {\bibinfo {author} {\bibfnamefont {C.-M.}\ \bibnamefont
  {Halati}}, \bibinfo {author} {\bibfnamefont {A.}~\bibnamefont {Sheikhan}}, \
  and\ \bibinfo {author} {\bibfnamefont {C.}~\bibnamefont {Kollath}},\
  }\href@noop {} {\bibfield  {journal} {\bibinfo  {journal} {arXiv preprint
  arXiv:2102.02537}\ } (\bibinfo {year} {2021})}\BibitemShut {NoStop}%
\bibitem [{\citenamefont {Carollo}\ and\ \citenamefont
  {Lesanovsky}(2021)}]{carollo2021nonequilibrium}%
  \BibitemOpen
  \bibfield  {author} {\bibinfo {author} {\bibfnamefont {F.}~\bibnamefont
  {Carollo}}\ and\ \bibinfo {author} {\bibfnamefont {I.}~\bibnamefont
  {Lesanovsky}},\ }\href@noop {} {\enquote {\bibinfo {title} {Nonequilibrium
  dark space phase transition},}\ } (\bibinfo {year} {2021}),\ \Eprint
  {http://arxiv.org/abs/2105.06729} {arXiv:2105.06729 [cond-mat.stat-mech]}
  \BibitemShut {NoStop}%
\bibitem [{\citenamefont {Plenio}\ and\ \citenamefont
  {Huelga}(2008)}]{Plenio_2008}%
  \BibitemOpen
  \bibfield  {author} {\bibinfo {author} {\bibfnamefont {M.~B.}\ \bibnamefont
  {Plenio}}\ and\ \bibinfo {author} {\bibfnamefont {S.~F.}\ \bibnamefont
  {Huelga}},\ }\href {\doibase 10.1088/1367-2630/10/11/113019} {\bibfield
  {journal} {\bibinfo  {journal} {New J. Phys.}\ }\textbf {\bibinfo {volume}
  {10}},\ \bibinfo {pages} {113019} (\bibinfo {year} {2008})}\BibitemShut
  {NoStop}%
\bibitem [{\citenamefont {Rebentrost}\ \emph {et~al.}(2009)\citenamefont
  {Rebentrost}, \citenamefont {Mohseni}, \citenamefont {Kassal}, \citenamefont
  {Lloyd},\ and\ \citenamefont {Aspuru-Guzik}}]{Rebentrost_2009}%
  \BibitemOpen
  \bibfield  {author} {\bibinfo {author} {\bibfnamefont {P.}~\bibnamefont
  {Rebentrost}}, \bibinfo {author} {\bibfnamefont {M.}~\bibnamefont {Mohseni}},
  \bibinfo {author} {\bibfnamefont {I.}~\bibnamefont {Kassal}}, \bibinfo
  {author} {\bibfnamefont {S.}~\bibnamefont {Lloyd}}, \ and\ \bibinfo {author}
  {\bibfnamefont {A.}~\bibnamefont {Aspuru-Guzik}},\ }\href {\doibase
  10.1088/1367-2630/11/3/033003} {\bibfield  {journal} {\bibinfo  {journal}
  {New J. Phys.}\ }\textbf {\bibinfo {volume} {11}},\ \bibinfo {pages} {033003}
  (\bibinfo {year} {2009})}\BibitemShut {NoStop}%
\bibitem [{\citenamefont {Mendoza-Arenas}\ \emph {et~al.}(2013)\citenamefont
  {Mendoza-Arenas}, \citenamefont {Grujic}, \citenamefont {Jaksch},\ and\
  \citenamefont {Clark}}]{jj2013prb}%
  \BibitemOpen
  \bibfield  {author} {\bibinfo {author} {\bibfnamefont {J.~J.}\ \bibnamefont
  {Mendoza-Arenas}}, \bibinfo {author} {\bibfnamefont {T.}~\bibnamefont
  {Grujic}}, \bibinfo {author} {\bibfnamefont {D.}~\bibnamefont {Jaksch}}, \
  and\ \bibinfo {author} {\bibfnamefont {S.~R.}\ \bibnamefont {Clark}},\ }\href
  {\doibase 10.1103/PhysRevB.87.235130} {\bibfield  {journal} {\bibinfo
  {journal} {Phys. Rev. B}\ }\textbf {\bibinfo {volume} {87}},\ \bibinfo
  {pages} {235130} (\bibinfo {year} {2013})}\BibitemShut {NoStop}%
\bibitem [{\citenamefont {Contreras-Pulido}\ \emph {et~al.}(2014)\citenamefont
  {Contreras-Pulido}, \citenamefont {Bruderer}, \citenamefont {Huelga},\ and\
  \citenamefont {Plenio}}]{Contreras_Pulido_2014}%
  \BibitemOpen
  \bibfield  {author} {\bibinfo {author} {\bibfnamefont {L.~D.}\ \bibnamefont
  {Contreras-Pulido}}, \bibinfo {author} {\bibfnamefont {M.}~\bibnamefont
  {Bruderer}}, \bibinfo {author} {\bibfnamefont {S.~F.}\ \bibnamefont
  {Huelga}}, \ and\ \bibinfo {author} {\bibfnamefont {M.~B.}\ \bibnamefont
  {Plenio}},\ }\href {\doibase 10.1088/1367-2630/16/11/113061} {\bibfield
  {journal} {\bibinfo  {journal} {New J. Phys.}\ }\textbf {\bibinfo {volume}
  {16}},\ \bibinfo {pages} {113061} (\bibinfo {year} {2014})}\BibitemShut
  {NoStop}%
\bibitem [{\citenamefont {Gorman}\ \emph {et~al.}(2018)\citenamefont {Gorman},
  \citenamefont {Hemmerling}, \citenamefont {Megidish}, \citenamefont
  {Moeller}, \citenamefont {Schindler}, \citenamefont {Sarovar},\ and\
  \citenamefont {Haeffner}}]{gorman2018prx}%
  \BibitemOpen
  \bibfield  {author} {\bibinfo {author} {\bibfnamefont {D.~J.}\ \bibnamefont
  {Gorman}}, \bibinfo {author} {\bibfnamefont {B.}~\bibnamefont {Hemmerling}},
  \bibinfo {author} {\bibfnamefont {E.}~\bibnamefont {Megidish}}, \bibinfo
  {author} {\bibfnamefont {S.~A.}\ \bibnamefont {Moeller}}, \bibinfo {author}
  {\bibfnamefont {P.}~\bibnamefont {Schindler}}, \bibinfo {author}
  {\bibfnamefont {M.}~\bibnamefont {Sarovar}}, \ and\ \bibinfo {author}
  {\bibfnamefont {H.}~\bibnamefont {Haeffner}},\ }\href {\doibase
  10.1103/PhysRevX.8.011038} {\bibfield  {journal} {\bibinfo  {journal} {Phys.
  Rev. X}\ }\textbf {\bibinfo {volume} {8}},\ \bibinfo {pages} {011038}
  (\bibinfo {year} {2018})}\BibitemShut {NoStop}%
\bibitem [{\citenamefont {Poto{\v{c}}nik}\ \emph {et~al.}(2018)\citenamefont
  {Poto{\v{c}}nik}, \citenamefont {Bargerbos}, \citenamefont {Schr\"oder},
  \citenamefont {Khan}, \citenamefont {Collodo}, \citenamefont {Gasparinetti},
  \citenamefont {Salath\'e}, \citenamefont {Creatore}, \citenamefont {Eichler},
  \citenamefont {T\"ureci}, \citenamefont {Chin},\ and\ \citenamefont
  {Wallraff}}]{potocnik2018nat}%
  \BibitemOpen
  \bibfield  {author} {\bibinfo {author} {\bibfnamefont {A.}~\bibnamefont
  {Poto{\v{c}}nik}}, \bibinfo {author} {\bibfnamefont {A.}~\bibnamefont
  {Bargerbos}}, \bibinfo {author} {\bibfnamefont {F.~A. Y.~N.}\ \bibnamefont
  {Schr\"oder}}, \bibinfo {author} {\bibfnamefont {S.~A.}\ \bibnamefont
  {Khan}}, \bibinfo {author} {\bibfnamefont {M.~C.}\ \bibnamefont {Collodo}},
  \bibinfo {author} {\bibfnamefont {S.}~\bibnamefont {Gasparinetti}}, \bibinfo
  {author} {\bibfnamefont {Y.}~\bibnamefont {Salath\'e}}, \bibinfo {author}
  {\bibfnamefont {C.}~\bibnamefont {Creatore}}, \bibinfo {author}
  {\bibfnamefont {C.}~\bibnamefont {Eichler}}, \bibinfo {author} {\bibfnamefont
  {H.~E.}\ \bibnamefont {T\"ureci}}, \bibinfo {author} {\bibfnamefont {A.~W.}\
  \bibnamefont {Chin}}, \ and\ \bibinfo {author} {\bibfnamefont
  {A.}~\bibnamefont {Wallraff}},\ }\href {\doibase 10.1038/s41467-018-03312-x}
  {\bibfield  {journal} {\bibinfo  {journal} {Nat. Commun.}\ }\textbf {\bibinfo
  {volume} {9}},\ \bibinfo {pages} {904} (\bibinfo {year} {2018})}\BibitemShut
  {NoStop}%
\bibitem [{\citenamefont {Maier}\ \emph {et~al.}(2019)\citenamefont {Maier},
  \citenamefont {Brydges}, \citenamefont {Jurcevic}, \citenamefont {Trautmann},
  \citenamefont {Hempel}, \citenamefont {Lanyon}, \citenamefont {Hauke},
  \citenamefont {Blatt},\ and\ \citenamefont {Roos}}]{maier2019prl}%
  \BibitemOpen
  \bibfield  {author} {\bibinfo {author} {\bibfnamefont {C.}~\bibnamefont
  {Maier}}, \bibinfo {author} {\bibfnamefont {T.}~\bibnamefont {Brydges}},
  \bibinfo {author} {\bibfnamefont {P.}~\bibnamefont {Jurcevic}}, \bibinfo
  {author} {\bibfnamefont {N.}~\bibnamefont {Trautmann}}, \bibinfo {author}
  {\bibfnamefont {C.}~\bibnamefont {Hempel}}, \bibinfo {author} {\bibfnamefont
  {B.~P.}\ \bibnamefont {Lanyon}}, \bibinfo {author} {\bibfnamefont
  {P.}~\bibnamefont {Hauke}}, \bibinfo {author} {\bibfnamefont
  {R.}~\bibnamefont {Blatt}}, \ and\ \bibinfo {author} {\bibfnamefont {C.~F.}\
  \bibnamefont {Roos}},\ }\href {\doibase 10.1103/PhysRevLett.122.050501}
  {\bibfield  {journal} {\bibinfo  {journal} {Phys. Rev. Lett.}\ }\textbf
  {\bibinfo {volume} {122}},\ \bibinfo {pages} {050501} (\bibinfo {year}
  {2019})}\BibitemShut {NoStop}%
\bibitem [{\citenamefont {Zerah-Harush}\ and\ \citenamefont
  {Dubi}(2020)}]{elinor2020prr}%
  \BibitemOpen
  \bibfield  {author} {\bibinfo {author} {\bibfnamefont {E.}~\bibnamefont
  {Zerah-Harush}}\ and\ \bibinfo {author} {\bibfnamefont {Y.}~\bibnamefont
  {Dubi}},\ }\href {\doibase 10.1103/PhysRevResearch.2.023294} {\bibfield
  {journal} {\bibinfo  {journal} {Phys. Rev. Research}\ }\textbf {\bibinfo
  {volume} {2}},\ \bibinfo {pages} {023294} (\bibinfo {year}
  {2020})}\BibitemShut {NoStop}%
\bibitem [{\citenamefont {Takasu}\ and\ \citenamefont
  {Takahashi}(2009)}]{YTakasu-JPSJ09}%
  \BibitemOpen
  \bibfield  {author} {\bibinfo {author} {\bibfnamefont {Y.}~\bibnamefont
  {Takasu}}\ and\ \bibinfo {author} {\bibfnamefont {Y.}~\bibnamefont
  {Takahashi}},\ }\href@noop {} {\bibfield  {journal} {\bibinfo  {journal} {J.
  Phys. Soc. Jpn.}\ }\textbf {\bibinfo {volume} {78}},\ \bibinfo {pages}
  {012001} (\bibinfo {year} {2009})}\BibitemShut {NoStop}%
\bibitem [{\citenamefont {Rubio-Abadal}\ \emph {et~al.}(2019)\citenamefont
  {Rubio-Abadal}, \citenamefont {Choi}, \citenamefont {Zeiher}, \citenamefont
  {Hollerith}, \citenamefont {Rui}, \citenamefont {Bloch},\ and\ \citenamefont
  {Gross}}]{rubio2019prx}%
  \BibitemOpen
  \bibfield  {author} {\bibinfo {author} {\bibfnamefont {A.}~\bibnamefont
  {Rubio-Abadal}}, \bibinfo {author} {\bibfnamefont {J.-y.}\ \bibnamefont
  {Choi}}, \bibinfo {author} {\bibfnamefont {J.}~\bibnamefont {Zeiher}},
  \bibinfo {author} {\bibfnamefont {S.}~\bibnamefont {Hollerith}}, \bibinfo
  {author} {\bibfnamefont {J.}~\bibnamefont {Rui}}, \bibinfo {author}
  {\bibfnamefont {I.}~\bibnamefont {Bloch}}, \ and\ \bibinfo {author}
  {\bibfnamefont {C.}~\bibnamefont {Gross}},\ }\href {\doibase
  10.1103/PhysRevX.9.041014} {\bibfield  {journal} {\bibinfo  {journal} {Phys.
  Rev. X}\ }\textbf {\bibinfo {volume} {9}},\ \bibinfo {pages} {041014}
  (\bibinfo {year} {2019})}\BibitemShut {NoStop}%
\bibitem [{\citenamefont {Vidal}(2004)}]{vidal2004prl}%
  \BibitemOpen
  \bibfield  {author} {\bibinfo {author} {\bibfnamefont {G.}~\bibnamefont
  {Vidal}},\ }\href {\doibase 10.1103/PhysRevLett.93.040502} {\bibfield
  {journal} {\bibinfo  {journal} {Phys. Rev. Lett.}\ }\textbf {\bibinfo
  {volume} {93}},\ \bibinfo {pages} {040502} (\bibinfo {year}
  {2004})}\BibitemShut {NoStop}%
\bibitem [{\citenamefont {Paeckel}\ \emph {et~al.}(2019)\citenamefont
  {Paeckel}, \citenamefont {K\"ohler}, \citenamefont {Swoboda}, \citenamefont
  {Manmana}, \citenamefont {Schollw\"ock},\ and\ \citenamefont
  {Hubig}}]{paeckel2019ann}%
  \BibitemOpen
  \bibfield  {author} {\bibinfo {author} {\bibfnamefont {S.}~\bibnamefont
  {Paeckel}}, \bibinfo {author} {\bibfnamefont {T.}~\bibnamefont {K\"ohler}},
  \bibinfo {author} {\bibfnamefont {A.}~\bibnamefont {Swoboda}}, \bibinfo
  {author} {\bibfnamefont {S.~R.}\ \bibnamefont {Manmana}}, \bibinfo {author}
  {\bibfnamefont {U.}~\bibnamefont {Schollw\"ock}}, \ and\ \bibinfo {author}
  {\bibfnamefont {C.}~\bibnamefont {Hubig}},\ }\href {\doibase
  https://doi.org/10.1016/j.aop.2019.167998} {\bibfield  {journal} {\bibinfo
  {journal} {Ann. Phys.}\ }\textbf {\bibinfo {volume} {411}},\ \bibinfo {pages}
  {167998} (\bibinfo {year} {2019})}\BibitemShut {NoStop}%
\bibitem [{\citenamefont {Mendoza-Arenas}\ \emph {et~al.}(2019)\citenamefont
  {Mendoza-Arenas}, \citenamefont {Rojas-Gamboa}, \citenamefont {Plenio},\ and\
  \citenamefont {Prior}}]{jj2019prb}%
  \BibitemOpen
  \bibfield  {author} {\bibinfo {author} {\bibfnamefont {J.~J.}\ \bibnamefont
  {Mendoza-Arenas}}, \bibinfo {author} {\bibfnamefont {D.~F.}\ \bibnamefont
  {Rojas-Gamboa}}, \bibinfo {author} {\bibfnamefont {M.~B.}\ \bibnamefont
  {Plenio}}, \ and\ \bibinfo {author} {\bibfnamefont {J.}~\bibnamefont
  {Prior}},\ }\href {\doibase 10.1103/PhysRevB.100.104307} {\bibfield
  {journal} {\bibinfo  {journal} {Phys. Rev. B}\ }\textbf {\bibinfo {volume}
  {100}},\ \bibinfo {pages} {104307} (\bibinfo {year} {2019})}\BibitemShut
  {NoStop}%
\bibitem [{\citenamefont {Stolpp}\ \emph {et~al.}(2020)\citenamefont {Stolpp},
  \citenamefont {K\"ohler}, \citenamefont {Manmana}, \citenamefont
  {Jeckelmann}, \citenamefont {Heidrich-Meisner},\ and\ \citenamefont
  {Paeckel}}]{stolpp2020arxiv}%
  \BibitemOpen
  \bibfield  {author} {\bibinfo {author} {\bibfnamefont {J.}~\bibnamefont
  {Stolpp}}, \bibinfo {author} {\bibfnamefont {T.}~\bibnamefont {K\"ohler}},
  \bibinfo {author} {\bibfnamefont {S.~R.}\ \bibnamefont {Manmana}}, \bibinfo
  {author} {\bibfnamefont {E.}~\bibnamefont {Jeckelmann}}, \bibinfo {author}
  {\bibfnamefont {F.}~\bibnamefont {Heidrich-Meisner}}, \ and\ \bibinfo
  {author} {\bibfnamefont {S.}~\bibnamefont {Paeckel}},\ }\href@noop {}
  {\enquote {\bibinfo {title} {{Comparative Study of State-of-the-Art
  Matrix-Product-State Methods for Lattice Models with Large Local Hilbert
  Spaces}},}\ } (\bibinfo {year} {2020}),\ \Eprint
  {http://arxiv.org/abs/2011.07412} {arXiv:2011.07412 [quant-ph]} \BibitemShut
  {NoStop}%
\bibitem [{Note1()}]{Note1}%
  \BibitemOpen
  \bibinfo {note} {Calculations were carried out using an optimized version of
  the Tensor Network Theory library \cite {tnt,tnt_review1} being developed by
  Paul Secular.}\BibitemShut {Stop}%
\bibitem [{\citenamefont {Essler}\ \emph {et~al.}(2005)\citenamefont {Essler},
  \citenamefont {Frahm}, \citenamefont {Göhmann}, \citenamefont {Klümper},\
  and\ \citenamefont {Korepin}}]{essler_frahm_gohmann_klumper_korepin_2005}%
  \BibitemOpen
  \bibfield  {author} {\bibinfo {author} {\bibfnamefont {F.~H.~L.}\
  \bibnamefont {Essler}}, \bibinfo {author} {\bibfnamefont {H.}~\bibnamefont
  {Frahm}}, \bibinfo {author} {\bibfnamefont {F.}~\bibnamefont {Göhmann}},
  \bibinfo {author} {\bibfnamefont {A.}~\bibnamefont {Klümper}}, \ and\
  \bibinfo {author} {\bibfnamefont {V.~E.}\ \bibnamefont {Korepin}},\ }\href
  {\doibase 10.1017/CBO9780511534843} {\emph {\bibinfo {title} {The
  One-Dimensional Hubbard Model}}}\ (\bibinfo  {publisher} {Cambridge
  University Press},\ \bibinfo {year} {2005})\BibitemShut {NoStop}%
\bibitem [{\citenamefont {Doyon}(2017)}]{BenjaminGGE}%
  \BibitemOpen
  \bibfield  {author} {\bibinfo {author} {\bibfnamefont {B.}~\bibnamefont
  {Doyon}},\ }\href {\doibase 10.1007/s00220-017-2836-7} {\bibfield  {journal}
  {\bibinfo  {journal} {Communications in Mathematical Physics}\ }\textbf
  {\bibinfo {volume} {351}},\ \bibinfo {pages} {155} (\bibinfo {year}
  {2017})}\BibitemShut {NoStop}%
\bibitem [{\citenamefont {Prelov\ifmmode~\check{s}\else \v{s}\fi{}ek}\ \emph
  {et~al.}(2016)\citenamefont {Prelov\ifmmode~\check{s}\else \v{s}\fi{}ek},
  \citenamefont {Bari\ifmmode \check{s}\else \v{s}\fi{}i\ifmmode~\acute{c}\else
  \'{c}\fi{}},\ and\ \citenamefont {\ifmmode \check{Z}\else
  \v{Z}\fi{}nidari\ifmmode~\check{c}\else \v{c}\fi{}}}]{prelovsek2016prb}%
  \BibitemOpen
  \bibfield  {author} {\bibinfo {author} {\bibfnamefont {P.}~\bibnamefont
  {Prelov\ifmmode~\check{s}\else \v{s}\fi{}ek}}, \bibinfo {author}
  {\bibfnamefont {O.~S.}\ \bibnamefont {Bari\ifmmode \check{s}\else
  \v{s}\fi{}i\ifmmode~\acute{c}\else \'{c}\fi{}}}, \ and\ \bibinfo {author}
  {\bibfnamefont {M.}~\bibnamefont {\ifmmode \check{Z}\else
  \v{Z}\fi{}nidari\ifmmode~\check{c}\else \v{c}\fi{}}},\ }\href {\doibase
  10.1103/PhysRevB.94.241104} {\bibfield  {journal} {\bibinfo  {journal} {Phys.
  Rev. B}\ }\textbf {\bibinfo {volume} {94}},\ \bibinfo {pages} {241104}
  (\bibinfo {year} {2016})}\BibitemShut {NoStop}%
\bibitem [{\citenamefont {{\v{Z}}nidari{\v{c}}}\ \emph
  {et~al.}(2017)\citenamefont {{\v{Z}}nidari{\v{c}}}, \citenamefont
  {Mendoza-Arenas}, \citenamefont {Clark},\ and\ \citenamefont
  {Goold}}]{vznidarivc2017dephasing}%
  \BibitemOpen
  \bibfield  {author} {\bibinfo {author} {\bibfnamefont {M.}~\bibnamefont
  {{\v{Z}}nidari{\v{c}}}}, \bibinfo {author} {\bibfnamefont {J.~J.}\
  \bibnamefont {Mendoza-Arenas}}, \bibinfo {author} {\bibfnamefont {S.~R.}\
  \bibnamefont {Clark}}, \ and\ \bibinfo {author} {\bibfnamefont
  {J.}~\bibnamefont {Goold}},\ }\href
  {https://onlinelibrary.wiley.com/doi/full/10.1002/andp.201600298} {\bibfield
  {journal} {\bibinfo  {journal} {Ann. Phys. (Berl.)}\ }\textbf {\bibinfo
  {volume} {529}},\ \bibinfo {pages} {1600298} (\bibinfo {year}
  {2017})}\BibitemShut {NoStop}%
\bibitem [{\citenamefont {L\"uschen}\ \emph {et~al.}(2017)\citenamefont
  {L\"uschen}, \citenamefont {Bordia}, \citenamefont {Hodgman}, \citenamefont
  {Schreiber}, \citenamefont {Sarkar}, \citenamefont {Daley}, \citenamefont
  {Fischer}, \citenamefont {Altman}, \citenamefont {Bloch},\ and\ \citenamefont
  {Schneider}}]{luschen2017prx}%
  \BibitemOpen
  \bibfield  {author} {\bibinfo {author} {\bibfnamefont {H.~P.}\ \bibnamefont
  {L\"uschen}}, \bibinfo {author} {\bibfnamefont {P.}~\bibnamefont {Bordia}},
  \bibinfo {author} {\bibfnamefont {S.~S.}\ \bibnamefont {Hodgman}}, \bibinfo
  {author} {\bibfnamefont {M.}~\bibnamefont {Schreiber}}, \bibinfo {author}
  {\bibfnamefont {S.}~\bibnamefont {Sarkar}}, \bibinfo {author} {\bibfnamefont
  {A.~J.}\ \bibnamefont {Daley}}, \bibinfo {author} {\bibfnamefont {M.~H.}\
  \bibnamefont {Fischer}}, \bibinfo {author} {\bibfnamefont {E.}~\bibnamefont
  {Altman}}, \bibinfo {author} {\bibfnamefont {I.}~\bibnamefont {Bloch}}, \
  and\ \bibinfo {author} {\bibfnamefont {U.}~\bibnamefont {Schneider}},\ }\href
  {\doibase 10.1103/PhysRevX.7.011034} {\bibfield  {journal} {\bibinfo
  {journal} {Phys. Rev. X}\ }\textbf {\bibinfo {volume} {7}},\ \bibinfo {pages}
  {011034} (\bibinfo {year} {2017})}\BibitemShut {NoStop}%
\bibitem [{\citenamefont {Gopalakrishnan}\ \emph {et~al.}(2017)\citenamefont
  {Gopalakrishnan}, \citenamefont {Islam},\ and\ \citenamefont
  {Knap}}]{gopalakrishnan2017prl}%
  \BibitemOpen
  \bibfield  {author} {\bibinfo {author} {\bibfnamefont {S.}~\bibnamefont
  {Gopalakrishnan}}, \bibinfo {author} {\bibfnamefont {K.~R.}\ \bibnamefont
  {Islam}}, \ and\ \bibinfo {author} {\bibfnamefont {M.}~\bibnamefont {Knap}},\
  }\href {\doibase 10.1103/PhysRevLett.119.046601} {\bibfield  {journal}
  {\bibinfo  {journal} {Phys. Rev. Lett.}\ }\textbf {\bibinfo {volume} {119}},\
  \bibinfo {pages} {046601} (\bibinfo {year} {2017})}\BibitemShut {NoStop}%
\bibitem [{\citenamefont {Nandkishore}(2015)}]{nandkishore2015prb}%
  \BibitemOpen
  \bibfield  {author} {\bibinfo {author} {\bibfnamefont {R.}~\bibnamefont
  {Nandkishore}},\ }\href {\doibase 10.1103/PhysRevB.92.245141} {\bibfield
  {journal} {\bibinfo  {journal} {Phys. Rev. B}\ }\textbf {\bibinfo {volume}
  {92}},\ \bibinfo {pages} {245141} (\bibinfo {year} {2015})}\BibitemShut
  {NoStop}%
\bibitem [{\citenamefont {Hyatt}\ \emph {et~al.}(2017)\citenamefont {Hyatt},
  \citenamefont {Garrison}, \citenamefont {Potter},\ and\ \citenamefont
  {Bauer}}]{hyatt2017prb}%
  \BibitemOpen
  \bibfield  {author} {\bibinfo {author} {\bibfnamefont {K.}~\bibnamefont
  {Hyatt}}, \bibinfo {author} {\bibfnamefont {J.~R.}\ \bibnamefont {Garrison}},
  \bibinfo {author} {\bibfnamefont {A.~C.}\ \bibnamefont {Potter}}, \ and\
  \bibinfo {author} {\bibfnamefont {B.}~\bibnamefont {Bauer}},\ }\href
  {\doibase 10.1103/PhysRevB.95.035132} {\bibfield  {journal} {\bibinfo
  {journal} {Phys. Rev. B}\ }\textbf {\bibinfo {volume} {95}},\ \bibinfo
  {pages} {035132} (\bibinfo {year} {2017})}\BibitemShut {NoStop}%
\end{thebibliography}%

%\newpage
%\onecolumngrid
%\newpage

%\renewcommand\thesection{S\arabic{section}}
%\renewcommand\theequation{S\arabic{equation}}
%\renewcommand\thefigure{S\arabic{figure}}
%\setcounter{equation}{0}

%\begin{center}
%{\Large \emph{Supplementary information}: Stochastic resonance in a quenched quantum many-body Bose-Fermi mixture}
%\end{center}

%\section{Details on the numerical simulation}
%We perform the time-evolving block decimation calculations using the open-source tensor network theory (TNT) library \cite{tnt,tnt_review1}.

\end{document}